\documentclass[11pt]{article}

\usepackage{amsmath,amssymb}
\usepackage{array}
\usepackage{color}

\usepackage{latexsym}

\newcount\figureno     \figureno=0
\newdimen\figdim       \figdim=70mm
\def\figureinc{%
   \global\advance\figureno by 1%
}
\def\figcaption#1#2#3{\hbox to #2{\hss{\vbox{\hsize=#2 \parindent=0pt
        {\bf Figure \number\figureno#3 :\ }#1}}\hss}
}

\begin{document}
\baselineskip 100pt

\large
\parskip.2in
\newcommand{\sdir}{\ensuremath{\rlap{\raisebox{0.15ex}{$\mskip 6.5mu\scriptstyle+$}}\supset}}
\newcommand{\acc}{\\[3mm]}

\def\mod#1{ \vert #1 \vert }
\def\chapter#1{\hbox{Introduction.}}
\newcommand{\R}{{\mathbb{R}}}
\nopagebreak[3]
\bigskip

\title{ \bf Invariant solutions of the supersymmetric sine--Gordon equation }
\vskip 1cm

\bigskip
\author{
A.~M. Grundland\thanks{email address: grundlan@crm.umontreal.ca}
\\
Centre de Recherches Math\'{e}matiques, Universit\'{e} de Montr\'{e}al,\\
C. P. 6128, Succ.\ Centre-ville, Montr\'{e}al, (QC) H3C 3J7,
Canada\\ Universit\'{e} du Qu\'{e}bec, Trois-Rivi\`{e}res, CP500 (QC) G9A 5H7, Canada \acc A. J. Hariton\thanks{email address: hariton@crm.umontreal.ca}
\\
Centre de Recherches Math\'{e}matiques, Universit\'{e} de Montr\'{e}al, \\
C. P. 6128, Succ.\ Centre-ville, Montr\'{e}al, (QC) H3C 3J7, Canada \acc L. \v{S}nobl\thanks{email address: Libor.Snobl@fjfi.cvut.cz}
\\
Faculty of Nuclear Sciences and Physical Engineering, \\
Czech Technical University in Prague, \\
B\v{r}ehov\'a 7, 115 19 Prague 1, Czech Republic} \date{}
\maketitle

\begin{abstract}

A comprehensive symmetry analysis of the ${\cal N}=1$ supersymmetric sine--Gordon equation is performed. Two different forms of the supersymmetric system are considered. We begin by studying a system of partial differential equations corresponding to the coefficients of the various powers of the anticommuting independent variables. Next, we consider the super-sine-Gordon equation expressed in terms of a bosonic superfield involving anticommuting independent variables.
In each case, a Lie (super)algebra of symmetries is determined and a classification of all subgroups having generic orbits of codimension 1 in the space of independent variables is performed. The method of symmetry reduction is systematically applied in order to derive invariant solutions of the supersymmetric model. Several types of algebraic, hyperbolic and doubly periodic solutions are obtained in explicit form. 
\end{abstract}

Short Title: Solutions of the SUSY sine--Gordon equation



\newpage

\section{Introduction}

The purpose of this paper is to obtain Lie point symmetries and group--invariant solutions of the minimal (${\cal N}=1$) supersymmetric extension of the $(1+1)$-dimensional sine--Gordon equation
\begin{equation}
\varphi_{xt}=\sin{\varphi}.
\label{intr1}
\end{equation}
The symmetry reduction method (SRM) is systematically applied in order to derive invariant solutions of the ${\cal N}=1$ supersymmetric extension of the model (\ref{intr1}).

 The classical sine--Gordon equation (\ref{intr1}) has applications in various areas of physics including, among others, nonlinear field theory, solid state physics (evolution of magnetic flux in Josephson junctions, Bloch wall motion of magnetic crystals, etc.), nonlinear optics (self--induced transparency, fiber optics), elementary particle theory and fluid dynamics; see \cite{Ablowitz1,Lamb,Scott,Barone,Bullough,Rogers} and references therein. A broad review of recent developments in the theory involved as well as their applications can be found for example in \cite{Lamb,Rogers,Ablowitz2,Novikov} and bibliographies therein. The sine--Gordon equation (\ref{intr1}) also has great significance in mathematics, especially in the soliton theory of surfaces. Analytic nonpertubative techniques for solving equation (\ref{intr1}) exist, including among others, the inverse scattering method and the Darboux--B\"{a}cklund transformations. Multiple soliton solutions of (\ref{intr1}) have found a wide variety of applications. The B\"{a}cklund transformation for the sine--Gordon equation (\ref{intr1}) linking different analytic descriptions of constant negative curvature surfaces in $\mathbb{R}^3$ was established a century ago by L. Bianchi \cite{Bianchi} and next by R. Steuerwald \cite{Steuerwald}. They were the first to find solutions of the structural equations (i.e. the Gauss--Weingarten and the Gauss--Codazzi--Mainardi equations). In particular, they constructed pseudospherical surfaces for the sine--Gordon equation (\ref{intr1}) by means of the auto--B\"{a}cklund transformation. It was demonstrated \cite{Ferapontov} that these surfaces can be described either by the Monge--Amp\`{e}re equation
\begin{equation}
u_{xx}u_{yy}-u_{xy}^2+\left(1+(u_x)^2+(u_y)^2\right)^2=0,
\label{intr2}
\end{equation}
(where $z=u(x,y)$ is the graph of a surface in $\mathbb{R}^3$) or by the sine--Gordon equation (\ref{intr1}) for the angle $\varphi(x,t)$ between asymptotic directions. The surfaces associated with equations (\ref{intr1}) and (\ref{intr2}) are characterized by the Gaussian curvature $K=-1$. The explicit form of the correspondence between these two integrable models is known \cite{Ferapontov}.

In recent publications (see e.g. \cite{DiVecchia,Chaichian,Grammaticos,Siddiq1,Siddiq2,Liu}), a superspace extension of the Lagrangian formulation has been established for the 
supersymmetric sine--Gordon (SSG) equation. The associated linear spectral problem was thoroughly discussed by many authors (see e.g. \cite{Chaichian,Liu} and references therein). It was shown \cite{Siddiq1} that the equation of motion appears as the compatibility condition of a set of Riccati equations. The supersymmetric sine--Gordon equation admits an infinite number of conservation laws, and a connection was established \cite{Siddiq1,Siddiq2} between its super--B\"{a}cklund and super--Darboux transformations. Consequently, it was shown in \cite{Siddiq2} that the Darboux transformation is related to the super--B\"{a}cklund transformation, and the latter was used to construct multi--super soliton solutions. The SSG equation was shown to be equivalent to the super $\mathbb{C}P^1$ sigma model \cite{Hruby,Witten}. The prolongation method of Wahlquist and Estabrook was used to find an infinite--dimensional superalgebra and the associated super Lax pairs \cite{Omote}.

In physics, the supersymmetric sine--Gordon equation is a useful example of a nonlinear integrable supersymmetric theory, on which conjectures concerning the  properties of such theories can be tested. These involve, among others, the computations of the S--matrix \cite{Ahn,Bajnok}. In addition, ${\cal N}=2$ supersymmetric sine--Gordon models arise in certain reductions of superstring worldsheet theories on particular backgrounds, e.g. the Pohlmeyer reduction on $AdS_2 \times S^2$ \cite{Grig}.

The purpose of this paper is to study the symmetry properties of the supersymmetric sine--Gordon system (more precisely, of the equations of motion of the ${\cal N}=1$ supersymmetric sine--Gordon model) and to construct various classes of invariant solutions of this model. In order to accomplish this, we use a generalized version of the prolongation method which encompasses commuting and anticommuting variables. The total derivatives with respect to these variables are adapted in such a way that they are consistent with the standard definitions. We then use a generalized version of the SRM in order to obtain group--invariant solutions of the supersymmetric sine--Gordon model. These solutions complement the multi--super soliton solutions found recently.

This paper is organized as follows. In Section \ref{sec2}, we recall the supersymmetric sine--Gordon equation, constructed in such a way that it is invariant under two independent supersymmetry transformations. In Section \ref{LsssGcf}, we decompose the supersymmetric sine--Gordon equation into three partial differential equations involving the component fields of the superfield and proceed to determine a Lie symmetry algebra of this system. Next, we focus on the SSG equation expressed explicitly in terms of the odd superspace variables $\theta_1$ and $\theta_2$ and the bosonic superfield $\Phi$. In Section \ref{sssGe}, we compute in detail the Lie superalgebra of symmetries of this equation using a generalized version of the prolongation method. The subalgebra classification of this superalgebra is performed in Section \ref{1dsaSSG}, and a discussion of the invariant solutions of the SSG equation is the subject of Section \ref{isSSG}. Finally, in Section \ref{concl}, we provide a summary of the results and list some possible future developments.

\section{Supersymmetric extension}\label{sec2}

We are interested in the supersymmetric sine--Gordon equation \cite{Grammaticos,Siddiq1,Siddiq2} constructed on the 4-dimensional superspace $\{(x,t,\theta_1,\theta_2)\}$. Here, $x$ and $t$ represent the even (bosonic) coordinates on the 2--dimensional super--Minkowski space $\R^{(1,1|2)} $, while the quantities $\theta_1$ and $\theta_2$ are anticommuting odd coordinates.

We replace the real--valued function $\varphi(x,t)$ in equation (\ref{intr1}) by the real scalar bosonic superfield $\Phi\left(x,t,\theta_1,\theta_2\right)$. Such a superfield can be decomposed into its component fields as
\begin{equation}
\Phi\left(x,t,\theta_1,\theta_2\right)=\frac{1}{2}u(x,t)+\theta_1\phi(x,t)+\theta_2\psi(x,t)+\theta_1\theta_2F(x,t),
\label{b1}
\end{equation}
where $\phi$ and $\psi$ are odd--valued functions (fields) and $u,F$ are even--valued functions (fields). 
The supersymmetric extension of the equation (\ref{intr1}) is constructed in such a way that it is invariant under the two independent supersymmetry transformations
\begin{equation}
x\rightarrow x-\underline{\eta}_1\theta_1,\quad \theta_1\rightarrow\theta_1+\underline{\eta}_1\qquad\mbox{ and }\qquad t\rightarrow t-\underline{\eta}_2\theta_2,\quad \theta_2\rightarrow\theta_2+\underline{\eta}_2,
\label{b2}
\end{equation}
where $\underline{\eta_1}$ and $\underline{\eta_2}$ are odd parameters (in general, we use the convention that underlined letters represent odd parameters). These transformations are generated by the infinitesimal supersymmetry generators
\begin{equation}
Q_x=\partial_{\theta_1}-\theta_1\partial_x\qquad\mbox{ and }\qquad Q_t=\partial_{\theta_2}-\theta_2\partial_t,
\label{b3}
\end{equation}
which satisfy the anticommutation relations
\begin{equation}
 \{Q_x,Q_x\}= -2\partial_x, \; \{Q_t,Q_t\}=-2\partial_t, \; \{Q_x,Q_t\}=0.
\end{equation}
In order to make our superfield theory invariant under the action $Q_x$ and $Q_t$, we write the equation in terms of the covariant derivatives 
\begin{equation}
D_x=\partial_{\theta_1}+\theta_1\partial_x\qquad\mbox{ and }\qquad D_t=\partial_{\theta_2}+\theta_2\partial_t,
\label{b4}
\end{equation}
which possess the property that they square to the generators of spacetime translations 
and anticommute with the supersymmetry generators
\begin{equation}
D_x^2=\partial_x,\quad D_t^2=\partial_t,\quad \{D_x,D_t\}=\{D_x,Q_x\}=\{D_x,Q_t\}=\{D_t,Q_x\}=\{D_t,Q_t\}=0.
\label{b5}
\end{equation}
The superspace Lagrangian density of the supersymmetric model is
\begin{equation}
{\mathcal L}(\Phi)=\frac{1}{2}D_x\Phi D_t\Phi-\cos{\Phi},
\label{b6}
\end{equation}
and the corresponding Euler--Lagrange superfield equation is given by
\begin{equation}
D_x D_t\Phi=\sin{\Phi}.
\label{b7}
\end{equation}
Equation (\ref{b7}) is invariant under the 
supersymmetry transformations (\ref{b2}), and we therefore refer to it as the supersymmetric sine-Gordon (SSG) equation. Once it is expanded out in terms of the component fields $\frac{1}{2}u(x,t),\phi(x,t),\psi(x,t),F(x,t)$ one finds that the scalar part of equation (\ref{b7}) is in fact algebraic and restricts $F$ to be the following function of $u$ \cite{Grammaticos}
\begin{equation}
F=-\sin{\left(\frac{u}{2}\right)}. 
\end{equation}
\medskip

Up to this point, the presentation has been formulated in the language usually used in physics, not yet mathematically well--defined. The mathematically sound formulation is based on the notion of supermanifolds in the sense of Refs. \cite{AR,CWMIS} and can be described as follows: 

One starts by considering a real Grassmann algebra $\Lambda$ generated by a finite or infinite number of generators $( \xi_1,\xi_2,\ldots )$. The number of Grassmann generators of $\Lambda$ is not directly relevant for applications, essentially the only assumption is ``there are at least as many independent ones as is needed in any formula encountered''. The Grassmann algebra $\Lambda$ has a naturally defined parity $\tilde 1=0, \tilde \xi_i=1,\widetilde{(ab)}={\tilde a}+{\tilde b} \ {\rm mod}\ 2$ and can be split into even and odd part
\begin{equation}
 \Lambda = \Lambda_{even}+\Lambda_{odd}.
\end{equation} 
The spaces $\Lambda$ and $\Lambda_{even}$ replace the field of real numbers in the context of supersymmetry. Elements of $\Lambda$ are called supernumbers, while elements of its even/odd part are even/odd supernumbers. For instance, in equation (\ref{b2}) we have parameters $\underline{\eta_1}$, $\underline{\eta_2}\in \Lambda_{odd}$. Sometimes we may also employ a different split
\begin{equation}\label{bodysoul}
 \Lambda = \Lambda_{body}+\Lambda_{soul}
\end{equation}
where $\Lambda_{body}=\wedge^{0}[\xi_1,\xi_2,\ldots]\simeq \R$ and $\Lambda_{soul}= \sum_{k\geq 1} \wedge^{k}[\xi_1,\xi_2,\ldots]$. The bodiless elements in $\Lambda_{soul}$ are obviously non--invertible because of the $\mathbb{Z}^{+}_{0}$--grading of the Grassmann algebra. If the number of Grassmann generators $K$ is finite, bodiless elements are nilpotent of degree at most $K$. In what follows we shall assume that $K$ is arbitrarily large but finite -- this assumption will allow us to use rigorous theorems of \cite{ARG}.

Next, one considers a $\mathbb{Z}_2$--graded real vector space $V$, with even basis elements $u_i$, $i=1,\ldots,N$ and odd basis elements $\upsilon_\mu$, $\mu=1,\ldots,M$ and constructs $W=\Lambda \otimes_{\R} V$. The space of interest to us is its even part
$$ W_{even} = \left\{  \sum_{i} a_i u_i +\sum_{\mu} \underline{\alpha}_\mu \upsilon_\mu | \ a_i \in \Lambda_{even}, \underline{\alpha}_\mu \in \Lambda_{odd} \right\}.$$
Obviously, $W_{even}$ is a $\Lambda_{even}$--module and can be identified with $ \Lambda_{even}^{\times N} \times \Lambda_{odd}^{\times M} $. To the original basis consisting of $u_i$ and $\upsilon_\mu$ (although $\upsilon_\mu \not\in W_{even}$ !) we associate the corresponding functionals 
$$ E_j: W_{even} \rightarrow \Lambda_{even}: \; E_j \left( \sum_{i} a_i v_i +\sum_{\mu} \underline{\alpha}_\mu \upsilon_\mu \right) = a_j, $$
$$ \Upsilon_\nu: W_{even} \rightarrow \Lambda_{odd}: \; \Upsilon_\nu \left( \sum_{i} a_i v_i +\sum_{\mu} \underline{\alpha}_\mu \upsilon_\mu \right) = \underline{\alpha}_\nu$$
and view them as the coordinates (even and odd, respectively) on $W_{even}$.  Any topological space locally diffeomorphic to a suitable $W_{even}$ is called a supermanifold.
The transition functions to even and odd coordinates between different charts on supermanifold are assumed to be even and odd--valued superanalytic or at least $G^{\infty}$ functions on $W_{even}$. For comprehensive definitions of the classes of ``supersmooth'' functions $G^{\infty}$ and superanalytic functions $G^{\omega}$ we refer the reader to consult e.g. the Ref. \cite{AR}, definition 2.5 -- here we only note that superanalytic functions are those that can be expanded into convergent power series in even and odd coordinates, whereas the definition of $G^{\infty}$ function is a more involved analogue on supermanifold of $C^{\infty}$ functions on manifolds. Any $G^{\infty}$ function can be expanded into products of odd coordinates (i.e. Taylor--like expansion) but the coefficients, being functions of even coordinates, may not necessarily be analytic.

In our context, the super--Minkowski space $\R^{(1,1|2)}$ should be understood as such a supermanifold, globally diffeomorphic to $ \Lambda_{even}^{\times 2} \times \Lambda_{odd}^{\times 2}$ with even coordinates $x,t$ and odd coordinates $\theta_1,\theta_2$. The supersymmetry transformation (\ref{b2}) can be viewed as a particular change of coordinates on $\R^{(1,1|2)}$ which transforms solutions of equation (\ref{b7}) into solutions of the same equation in new coordinates.

A bosonic, also called even, superfield is a $G^{\infty}$ function
$ \Phi: \R^{(1,1|2)} \rightarrow \Lambda_{even}.$
It can be expanded in powers of odd coordinates $\theta_1,\theta_2$ giving the decomposition (\ref{b1}), with 
$$u, \, F : \Lambda_{even}^{\times 2} \rightarrow \Lambda_{even}, $$
$$\phi, \, \psi : \Lambda_{even}^{\times 2} \rightarrow \Lambda_{odd}.$$

The partial derivatives with respect to odd coordinate (for a detailed description see \cite{AR}, definitions 2.5 and 5.6) satisfy the usual operational rules, namely $\partial_{\theta_i} \theta_j=\delta^i_j$ together with the graded product rule 
\begin{equation}
\label{gLr}
\partial_{\theta_i} (fg) = (\partial_{\theta_i} f)g+(-1)^{\tilde f} (\partial_{\theta_i} g).
\end{equation}
The operations $\partial_{\theta_i},Q_{x,t},D_{x,t}$  in equations (\ref{b3}) and (\ref{b4}) switch the parity of the function acted on. For instance $\partial_{\theta_1} \Phi$ becomes an odd superfield $\partial_{\theta_1} \Phi: \R^{(1,1|2)} \rightarrow \Lambda_{odd}$ whose component decomposition is
$$ \partial_{\theta_1} \Phi\left(x,t,\theta_1,\theta_2\right)= \phi(x,t)+\theta_2 F(x,t).$$

\section{Lie symmetry properties of the supersymmetric sine--Gordon system in component form}\label{LsssGcf}

When decomposed in terms of the various powers of $\theta_1$ and $\theta_2$, the SSG equation (\ref{b7}) is seen to be equivalent to a system of three partial differential equations for the fields $u$, $\phi$ and $\psi$. That is, the coefficients of the powers $\theta_1$, $\theta_2$ and $\theta_1\theta_2$ combine to form
the following system of coupled equations for the component fields \cite{Grammaticos}
\begin{equation}
\begin{split}
&\mbox{(i) }u_{xt}=-\sin{u} + 2
\phi\psi\sin{\left(\frac{u}{2}\right)},\\ &\mbox{(ii) }\phi_t=-\psi\cos{\left(\frac{u}{2}\right)},\\ &\mbox{(iii) }\psi_x=\phi\cos{\left(\frac{u}{2}\right)}.
\end{split}
\label{c1}
\end{equation}
In order to determine the Lie point symmetry algebra $\mathfrak{g}$ of the system (\ref{c1}), we restrict our consideration to Lie groups and use an infinitesimal approach. We adapt the method of prolongation of vector fields described in the book by P. J. Olver \cite{Olver} to the case where the equations of the system contain both even and odd--valued functions \cite{HusAyaWin,HusAlvarez}. We begin by writing the set of partial differential equations (\ref{c1}) in the form 
\begin{equation}
\Delta_k(x,t,u,\phi,\psi,u_{xt},\phi_t,\psi_x)=0,\qquad k=1,2,3,
\label{c0001}
\end{equation}
where
\begin{equation}
\begin{split}
&\Delta_1(x,t,u,\phi,\psi,u_{xt},\phi_t,\psi_x)=u_{xt}+\sin{u}-2\phi\psi\sin{\left(\frac{u}{2}\right)},\\
&\Delta_2(x,t,u,\phi,\psi,u_{xt},\phi_t,\psi_x)=\phi_t+\psi\cos{\left(\frac{u}{2}\right)},\\
&\Delta_3(x,t,u,\phi,\psi,u_{xt},\phi_t,\psi_x)=\psi_x-\phi\cos{\left(\frac{u}{2}\right)},
\end{split}
\label{c0002}
\end{equation}

A symmetry group $G$ of the system (\ref{c0001}) is a (local) group of transformations acting on the cartesian product of supermanifolds 
$$
X\times U 
$$
with even coordinates $(x,t,u)$ and odd coordinates $(\phi,\psi)$,
whose associated action on the functions $u(x,t),\Phi(x,t),\psi(x,t)$ maps solutions of (\ref{c0001}) to solutions of (\ref{c0001}). Assuming that $G$ is a super Lie group in the sense of \cite{ARG} one can associate to it its Lie algebra of even left--invariant vector fields ${\mathcal G}$, whose elements are the infinitesimal symmetries of the system (\ref{c0001}). In particular, a local one--parameter subgroup of $G$ consists of a family of transformations
\begin{equation}
g_{\varepsilon}:\quad \tilde{x}^i = X^i(x,u,\varepsilon),\quad \tilde{u}^{\alpha} = U^{\alpha}(x,u,\varepsilon)
\end{equation}
where $x=(x^1,x^2)=(x,t)$ are the independent variables and $u=(u^1,u^2,u^3)=(u,\phi,\psi)$ the dependent variables. $\varepsilon\in\Lambda_{even}$ is a group parameter, whose range may be restricted depending on the values of $x,t,u,\phi,\psi$. Such a local subgroup is generated by a vector field of the form
\begin{equation}
\mathbf{v} = \xi^i(x,u)\frac{\partial}{\partial x^i} + \Phi(x,u)^{\alpha}\frac{\partial}{\partial u^{\alpha}}
\label{thisfield1}
\end{equation}
where
\begin{equation}
\xi^i(x,u) = \frac{\partial}{\partial\varepsilon}X^i|_{\varepsilon = 0},\quad \Phi^{\alpha}(x,u) = \frac{\partial}{\partial\varepsilon}U^{\alpha}|_{\varepsilon = 0}.
\end{equation}
The advantage of working with the Lie algebra $\mathfrak{g}$ instead of directly with the  super Lie group $G$ is that {the equations defining} the infinitesimal symmetries are linear.

In order to determine the infinitesimal symmetries of a system of partial differential equations, it is useful to make use of the concept of  the prolongation of a group action. The idea is that a transformation  of coordinates $x^i\rightarrow\tilde{x}^i$, $u^{\alpha}\rightarrow\tilde{u}^{\alpha}$ induces a transformation of the derivatives
\begin{equation}
\frac{\partial u^{\alpha}}{\partial x^i}\longrightarrow \frac{\partial \tilde{u}^{\alpha}}{\partial \tilde{x}^i}.
\end{equation}
In order to make use of this concept, we define the multi--index $J = (j_1,\ldots,j_p)$, where $j_i = 0,1,\ldots$ and
$|J| = j_1 + \ldots + j_p$. 
The space of coordinates on $X\times U$ is extended to the jet bundle
\begin{equation}
{\mathcal J}_k = \{(x^i,u^{\alpha},u^{\alpha}_J)| \, |J|\leq k\}
\end{equation}
which includes the coordinates and all derivatives of the dependent variables of order less than or equal to $k$.  In our setting the jet bundle is a supermanifold as well, since $X\times U$ was. On the jet bundle, we define total derivatives
\begin{equation} {\mathcal D}_i = \frac{\partial}{\partial x^i} +
  \sum\limits_{\alpha,J}u^{\alpha}_{J_i}\frac{\partial}{\partial
    u^{\alpha}_J}\mbox{,}\end{equation}
where $J_i = (j_1,\ldots,j_{i-1},j_i+1,j_{i+1},\ldots,j_n)$.
More generally, for $J = (j_1,j_2,\ldots,j_n)$, we define
\begin{equation}{\mathcal D}_J = \underbrace{{\mathcal D}_1{\mathcal D}_1\cdots {\mathcal D}_1}_{j_1} \cdots\ \underbrace{{\mathcal D}_n{\mathcal D}_n\cdots {\mathcal D}_n}_{j_n}. \end{equation} 

The prolongation of a group action to the jet bundle ${\mathcal J}_k$ in turn induces a prolongation of the generating infinitesimal vector field in the Lie algebra. For the vector field $\mathbf{v}$ given by (\ref{thisfield1}), the $k^{th}$ order prolongation of $\mathbf{v}$ is
\begin{equation}pr^{(k)}(\mathbf{v}) = \mathbf{v} + \sum\limits_{\alpha,|J|\neq 0}\phi^{\alpha}_J(x,u^{(k)})\frac{\partial}{\partial u^{\alpha}_J}\mbox{,} \end{equation}
where  $\phi^{\alpha}_J(x,u^{(k)})$ are given by the formula
\begin{equation}\label{xprolong}\phi^{\alpha}_J = {\mathcal D}_J\left(\phi^{\alpha} - \xi^i\frac{\partial u^{\alpha}}{\partial x^i}\right) + \xi^i u^{\alpha}_{J_i}
\end{equation}
or, equivalently, by the recursive formula
\begin{equation}\label{prbos}
 \phi^{\alpha}_{J_j}={\mathcal D}_{j} \phi^\alpha_J - \sum_{i} ({\mathcal D}_{j} \xi^i) u^{\alpha}_{J_i}.
\end{equation}

The symmetry criterion (Theorem 2.31 in \cite{Olver})  assumes that $G$ is a connected Lie group of transformations acting locally on $X\times
  U$ through the transformations
\begin{displaymath}\tilde{x}_i = X^i(x,u,g), \qquad \tilde{u}^{\alpha} = U^{\alpha}(x,u,g)\mbox{,}\end{displaymath}
where $g\in G$ and $\Delta_{\nu}(x,u^{(n)})$ is a non--degenerate system of partial differential equations (meaning that the system is locally solvable with respect to highest derivatives and is of maximal rank at every point $(x_0,u_0^{(n)})\in X\times U^{(n)}$). Then $G$ is a symmetry group of $\Delta = 0$ if and only if
\begin{equation}\left[pr^{(k)}(\mathbf{v})\right](\Delta) = 0 \  \mbox{whenever} \  \Delta = 0\mbox{,}\end{equation}
for each infinitesimal generator $\mathbf{v}$ of $G$.

Using the results of \cite{ARG}, one finds that the same criterion can be used also in the case of the super Lie group $G$ and its Lie algebra of even left--invariant vector fields.

For the purpose of determining the Lie algebra of symmetries of the system (\ref{c0001}), let us write a vector field of the form
\begin{equation}
\begin{split}
\mathbf{v}=&\xi(x,t,u,\phi,\psi)\partial_x+\tau(x,t,u,\phi,\psi)\partial_t+{\mathcal U}(x,t,u,\phi,\psi)\partial_u\\ &+\Sigma(x,t,u,\phi,\psi)\partial_{\phi}+\Psi(x,t,u,\phi,\psi)\partial_{\psi},
\end{split}
\label{c1A}
\end{equation}
where $\xi$, $\tau$ and ${\mathcal U}$ are  $\Lambda_{even}$--valued functions
while $\Sigma$ and $\Psi$ are  $\Lambda_{odd}$--valued so that $\mathbf{v}$ is an even vector field. 
We consider a 2nd prolongation of the vector field (\ref{c1A}) which is of the form
\begin{equation}
\mbox{pr}^{(2)}(\mathbf{v})=\mathbf{v}+{\mathcal U}^{xt}\partial_{u_{xt}}+\Sigma^t\partial_{\phi_t}+\Psi^x\partial_{\psi_x}+({\mathcal U}^x\partial_{u_x}+{\mathcal U}^t\partial_{u_t}+\ldots),
\label{c1B}
\end{equation}
where the terms in the parentheses don't contribute in what follows, namely in equation (\ref{c1E}). The coefficients ${\mathcal U}^{xt}$, $\Sigma^t$ and $\Psi^x$ are known functions of the components $\xi,\ldots,\Psi$ of the vector field $\mathbf{v}$ and their derivatives with respect to the independent and dependent variables $x,\ldots,\psi$ (as given by the general prolongation formula (\ref{xprolong}) or (\ref{prbos})). We use upper indices in coefficients ${\mathcal U}^{xt}$, $\Sigma^t$ etc. in order to distinguish them from partial derivatives, e.g. ${\mathcal U}_{xt} = \partial_{t}\partial_{x} {\mathcal U}$. The first order coefficients are given by
\begin{equation}
\Sigma^t=\Sigma_t+\Sigma_uu_t+\Sigma_{\phi}\phi_t+\Sigma_{\psi}\psi_t-\xi_t\phi_x-\xi_uu_t\phi_x-\xi_{\phi}\phi_x\phi_t-\xi_{\psi}\phi_x\psi_t-\tau_t\phi_t-\tau_uu_t\phi_t-\tau_{\psi}\phi_t\psi_t,
\label{c1C}
\end{equation}
and
\begin{equation}
\Psi^x=\Psi_x+\Psi_uu_x+\Psi_{\phi}\phi_x+\Psi_{\psi}\psi_x-\xi_x\psi_x-\xi_uu_x\psi_x+\xi_{\phi}\phi_x\psi_x-\tau_x\psi_t-\tau_uu_x\psi_t+\tau_{\phi}\phi_x\psi_t+\tau_{\psi}\psi_x\psi_t,
\label{c1D}
\end{equation}
The second order coefficient, ${\mathcal U}^{xt}$, is very involved and will not be presented here.

According to the symmetry criterion, the vector field (\ref{c1A}) is an infinitesimal generator of the symmetry group of the system of differential equations (\ref{c0001}) if and only if
\begin{equation}
\mbox{pr}^{(2)}(\mathbf{v})\left[\Delta_k(x,t,u,\phi,\psi,u_{xt},\phi_t,\psi_x)\right]=0,\quad k=1,2,3,
\label{c1E}
\end{equation}
whenever $\Delta_l(x,t,u,\phi,\psi,u_{xt},\phi_t,\psi_x)=0$, $l=1,2,3$.

The condition $\mbox{pr}^{(2)}(\mathbf{v})[\Delta_k]=0$, when applied to the system (\ref{c1}), leads to the following conditions on the coefficients
\begin{equation}
\begin{split}
&\mbox{(i) }{\mathcal U}^{xt}={\mathcal U}\left(-\cos{u}+\cos{\left(\frac{u}{2}\right)}\phi\psi\right)+\Sigma\left(2\sin{\left(\frac{u}{2}\right)}\psi\right)+\Psi\left(-2\sin{\left(\frac{u}{2}\right)}\phi\right),\\ &\mbox{(ii) }\Sigma^t=\frac{1}{2}{\mathcal U}\sin{\left(\frac{u}{2}\right)}\psi-\Psi\cos{\left(\frac{u}{2}\right)},\\ &\mbox{(iii) }\Psi^x=-\frac{1}{2}{\mathcal U}\sin{\left(\frac{u}{2}\right)}\phi+\Sigma\cos{\left(\frac{u}{2}\right)},
\end{split}
\label{c1F}
\end{equation}
whenever $u$, $\phi$, $\psi$ satisfy the system (\ref{c1}).

Substituting the prolongation formulae for ${\mathcal U}^{xt}$, $\Sigma^t$ and $\Psi^x$ into (\ref{c1F}) and imposing the condition that $\Delta_k=0$, $k=1,2,3$, i.e. substituting for $u_{xt}$, $\phi_t$, $\psi_x$ and their derivatives, we equate the coefficients of the various monomials in the various remaining derivatives of $u$, $\phi$ and $\psi$ with respect to $x$ and $t$ (i.e. those unconstrained by equation (\ref{c1})). We obtain a series of determining equations which impose restrictions on the coefficients $\xi$, $\tau$, ${\mathcal U}$, $\Sigma$ and $\Psi$ of the vector field (\ref{c1A}). Solving the determining equations, we see that
the coefficients must be
\begin{equation}
\xi(x)=C_1x+C_2,\quad \tau(t)=-C_1t+C_3,\quad {\mathcal U}=0,\quad \Sigma(\phi)=-\frac{1}{2}C_1\phi,\quad \Psi(\psi)=\frac{1}{2}C_1\psi,
\label{c1G}
\end{equation}
where $C_1$, $C_2$ and $C_3$ are arbitrary even parameters. Thus, we have determined that the Lie algebra $\mathfrak{g}$ is spanned by the following three vector fields
\begin{equation}
P_x=\partial_x,\qquad P_t=\partial_t,\qquad D=2x\partial_x-2t\partial_t-\phi\partial_{\phi}+\psi\partial_{\psi},
\label{c2}
\end{equation}
where $\partial_x=\partial/\partial x$ etc. We have two translations, $P_x$ and $P_t$, in the $x$ and $t$ directions respectively, and the dilation $D$ acting on the independent and dependent variables. We note that although the method may in general yield a super Lie algebra (for explicit examples see e.g. \cite{Hariton4,Hariton8,Hariton9}), in our particular case of the supersymmetric sine--Gordon system (\ref{c1}) the result is just a Lie algebra acting on the {supermanifold $X\times U$}. In fact, the Lie algebra in question whose nonzero commutation relations are
\begin{equation}
[P_x,D]=2P_x,\qquad [P_t,D]=-2P_t,
\label{c4}
\end{equation}
is $ISO(1,1)$, which is also the symmetry Lie algebra of the ordinary sine--Gordon equation (in $(1+1)$-dimensional  Minkowski space described in the light-cone coordinates, the  Lorentz boost takes the form of a dilation).
This represents the Poincar\'{e} invariance of the sine--Gordon equation, supersymmetric or otherwise. This algebra is also identified as  $A_{3,4}\left(E(1,1)\right)$ in \cite{Patera} where its  non--conjugate one--dimensional subalgebras are found to be
\begin{equation}
L_1=\{D\},\quad L_2=\{P_x\},\quad L_3=\{P_t\},\quad L_4=\{P_x+P_t\},\quad L_5=\{P_x-P_t\}.
\label{c3}
\end{equation}

One can now proceed to apply the SRM in order to obtain invariant solutions of the supersymmetric system (\ref{c1}). First, we find for each of the subalgebras listed in (\ref{c3}) the associated four invariants along with the appropriate change of variable that has to be substituted into the system (\ref{c1}) in order to obtain the set of reduced ordinary differential equations. In each case, the invariant involving only the independent variables, the so--called symmetry variable, is labelled by the symbol $\sigma$. The invariants and the change of variables are listed in Table 1, while the systems of reduced ordinary differential equations are listed in Table \ref{tb2}. Because the reduced ODE systems in Table \ref{tb2} form a subset of cases investigated in Section \ref{isSSG} we postpone their discussion there.

\begin{table}[htbp]
  \begin{center}
\caption{Invariants and change of variables for subalgebras of the Lie algebra $\mathfrak{g}$ spanned by the
  vector fields (\ref{c2})}
\vspace{3mm}
\setlength{\extrarowheight}{4pt}
\begin{tabular}{|c|c|c|}\hline
Subalgebra & Invariants & Relations and
Change of Variable\\[0.5ex]\hline\hline
$L_1=\{D\}$ & $\sigma=xt$, $u$, $t^{-1/2}\phi$, $t^{1/2}\psi$  & $u=u(\sigma)$, $\phi=t^{1/2}\varTheta(\sigma)$, $\psi=t^{-1/2}\Omega(\sigma)$, \\\hline
$L_2=\{P_x\}$ & $\sigma=t$, $u$, $\phi$, $\psi$  & $u=u(t)$, $\phi=\phi(t)$, $\psi=\psi(t)$, \\\hline
$L_3=\{P_t\}$ & $\sigma=x$, $u$, $\phi$, $\psi$  & $u=u(x)$, $\phi=\phi(x)$, $\psi=\psi(x)$, \\\hline
$L_4=\{P_x+P_t\}$ & $\sigma=x-t$, $u$, $\phi$, $\psi$  & $u=u(\sigma)$, $\phi=\phi(\sigma)$, $\psi=\psi(\sigma)$, \\\hline
$L_5=\{P_x-P_t\}$ & $\sigma=x+t$, $u$, $\phi$, $\psi$  & $u=u(\sigma)$, $\phi=\phi(\sigma)$, $\psi=\psi(\sigma)$, \\\hline
\end{tabular}
  \end{center}
\end{table}

\begin{table}[htbp]
  \begin{center}
\caption{Reduced equations obtained for subalgebras of the Lie algebra $\mathfrak{g}$ spanned by the
  vector fields (\ref{c2})}
\setlength{\extrarowheight}{4pt}
\begin{tabular}{|c|c|}\hline
Subalgebra & Reduced equations \\[0.5ex]\hline\hline
$L_1=\{D\}$ & $\sigma u_{\sigma\sigma}+u_{\sigma}=-\sin{u}+2\sin{\left(\frac{u}{2}\right)}\varTheta\Omega$,\qquad $\frac{1}{ 2}\varTheta+\sigma\varTheta_{\sigma}=-\cos{\left(\frac{u}{2}\right)}\Omega$,\qquad $\Omega_{\sigma}=\cos{\left(\frac{u}{2}\right)}\varTheta$ \\\hline
$L_2=\{P_x\}$ & $-\sin{u}+2\sin{\left(\frac{u}{2}\right)}\phi\psi=0$,\qquad $\phi_t=-\cos{\left(\frac{u}{2}\right)}\psi$,\qquad $\cos{\left(\frac{u}{2}\right)}\phi=0$ \\\hline
$L_3=\{P_t\}$ & $-\sin{u}+2\sin{\left(\frac{u}{2}\right)}\phi\psi=0$,\qquad $\cos{\left(\frac{u}{2}\right)}\psi=0$,\qquad $\psi_x=\cos{\left(\frac{u}{2}\right)}\phi$ \\\hline
$L_4=\{P_x+P_t\}$ & $-u_{\sigma\sigma}=-\sin{u}+2\sin{\left(\frac{u}{2}\right)}\phi\psi$,\qquad $\phi_{\sigma}=\cos{\left(\frac{u}{2}\right)}\psi$,\qquad $\psi_{\sigma}=\cos{\left(\frac{u}{2}\right)}\phi$ \\\hline
$L_5=\{P_x-P_t\}$ & $u_{\sigma\sigma}=-\sin{u}+2\sin{\left(\frac{u}{2}\right)}\phi\psi$,\qquad $\phi_{\sigma}=-\cos{\left(\frac{u}{2}\right)}\psi$,\qquad $\psi_{\sigma}=\cos{\left(\frac{u}{2}\right)}\phi$ \\\hline
\end{tabular}
  \end{center}\label{tb2}
\end{table}

\section{Symmetries of the SSG equation}\label{sssGe}

When we performed the group--theoretical analysis of the supersymmetric sine--Gordon system in the component form (\ref{c1}), we noted that the resulting symmetry algebra did not essentially differ from the purely bosonic case, i.e. equation (\ref{intr1}). That is, the supersymmetry algebra was not recovered in this way. In order to overcome this shortcoming of the method, we now turn our attention to the superfield version of the model represented by the SSG equation (\ref{b7}). This equation
can be re--written in the form
\begin{equation}
\theta_1\theta_2\Phi_{xt}-\theta_2\Phi_{t\theta_1}+\theta_1\Phi_{x\theta_2}-\Phi_{\theta_1\theta_2}=\sin{\Phi},
\label{symmie2}
\end{equation}
where each successive subscript (from left to right) indicates a successive partial derivative (for example $\Phi_{\theta_1\theta_2}$ represents $\partial_{\theta_2}\left(\partial_{\theta_1}\Phi\right)$). In order to determine the Lie superalgebra of symmetries of equation (\ref{symmie2}), we employ the generalized method of prolongations so as to include also the two independent odd variables $\theta_1$ and $\theta_2$.  Such procedure was proposed and used in \cite{AyaHus,AyaAyaHus}. 

 We consider transformations on the supermanifold
$$
\R^{(1,1|2)} \times \Lambda_{even}.
$$
We write a generator of symmetry transformation in the form of an even vector field on this manifold
\begin{equation}
\begin{split}
\mathbf{v}=&\xi(x,t,\theta_1,\theta_2,\Phi)\partial_x+\tau(x,t,\theta_1,\theta_2,\Phi)\partial_t+\rho(x,t,\theta_1,\theta_2,\Phi)\partial_{\theta_1}\\ &+\sigma(x,t,\theta_1,\theta_2,\Phi)\partial_{\theta_2}+\Pi(x,t,\theta_1,\theta_2,\Phi)\partial_{\Phi},
\end{split}
\label{symmie3}
\end{equation}
where $\xi$, $\tau$ and $\Pi$ are supposed to be even, i.e. $\Lambda_{even}$--valued functions, while $\rho$ and $\sigma$ are odd, i.e. $\Lambda_{odd}$--valued. Here, we adopt the ordering convention that the odd coefficients in the expression (in this case $\rho$ and $\sigma$) precede the odd derivatives ($\partial_{\theta_1}$ and $\partial_{\theta_2}$ respectively). 
We generalize the total derivatives ${\mathcal D}_x$, ${\mathcal D}_t$, ${\mathcal D}_{\theta_1}$ and ${\mathcal D}_{\theta_2}$ as
\begin{equation}
\begin{split}
{\mathcal D}_x=&\partial_x+\Phi_x\partial_{\Phi}+\Phi_{xx}\partial_{\Phi_x}+\Phi_{xt}\partial_{\Phi_t}+\Phi_{x\theta_1}\partial_{\Phi_{\theta_1}}+\Phi_{x\theta_2}\partial_{\Phi_{\theta_2}}+\Phi_{xxx}\partial_{\Phi_{xx}}+\Phi_{xxt}\partial_{\Phi_{xt}}\\ &+\Phi_{xx\theta_1}\partial_{\Phi_{x\theta_1}}+\Phi_{xx\theta_2}\partial_{\Phi_{x\theta_2}}+\Phi_{xtt}\partial_{\Phi_{tt}}+\Phi_{xt\theta_1}\partial_{\Phi_{t\theta_1}}+\Phi_{xt\theta_2}\partial_{\Phi_{t\theta_2}}+\Phi_{x\theta_1\theta_2}\partial_{\Phi_{\theta_1\theta_2}},
\end{split}
\label{symmie4}
\end{equation}
and
\begin{equation}
\begin{split}
{\mathcal D}_{\theta_1}=&\partial_{\theta_1}+\Phi_{\theta_1}\partial_{\Phi}+\Phi_{x\theta_1}\partial_{\Phi_x}+\Phi_{t\theta_1}\partial_{\Phi_t}+\Phi_{\theta_2\theta_1}\partial_{\Phi_{\theta_2}}+\Phi_{xx\theta_1}\partial_{\Phi_{xx}}+\Phi_{xt\theta_1}\partial_{\Phi_{xt}}\\ &+\Phi_{x\theta_2\theta_1}\partial_{\Phi_{x\theta_2}}+\Phi_{tt\theta_1}\partial_{\Phi_{tt}}+\Phi_{t\theta_2\theta_1}\partial_{\Phi_{t\theta_2}},
\end{split}
\label{symmie5}
\end{equation}
while ${\mathcal D}_t$ and ${\mathcal D}_{\theta_2}$ are defined in analogy with ${\mathcal D}_x$ and ${\mathcal D}_{\theta_1}$ respectively. Here, we note that the chain rule for an odd--valued function $f(g(x))$ is \cite{Berezin,DeWitt}
\begin{equation}
{\partial f\over \partial x}={\partial g\over \partial x}\cdot{\partial f\over \partial g}.
\label{chainrule}
\end{equation}
The graded interchangeability of mixed derivatives (i.e. with proper respect to the ordering of odd variables) of course holds. The second prolongation of the vector field (\ref{symmie3}) is given by
\begin{equation}
\begin{split}
\mbox{pr}^{(2)}\mathbf{v}=&\xi\partial_x+\tau\partial_t+\rho\partial_{\theta_1}+\sigma\partial_{\theta_2}+\Pi\partial_{\Phi}+\Pi^x\partial_{\Phi_x}+\Pi^t\partial_{\Phi_t}+\Pi^{\theta_1}\partial_{\Phi_{\theta_1}}+\Pi^{\theta_2}\partial_{\Phi_{\theta_2}}\\ &+\Pi^{xx}\partial_{\Phi_{xx}}+\Pi^{xt}\partial_{\Phi_{xt}}+\Pi^{x\theta_1}\partial_{\Phi_{x\theta_1}}+\Pi^{x\theta_2}\partial_{\Phi_{x\theta_2}}+\Pi^{tt}\partial_{\Phi_{tt}}+\Pi^{t\theta_1}\partial_{\Phi_{t\theta_1}}\\ &+\Pi^{t\theta_2}\partial_{\Phi_{t\theta_2}}+\Pi^{\theta_1\theta_2}\partial_{\Phi_{\theta_1\theta_2}}.
\end{split}
\label{symmie6}
\end{equation}
Applying the second prolongation (\ref{symmie6}) to the equation (\ref{symmie2}), we obtain the following condition
\begin{equation}
\begin{split}
&\rho\left(\theta_2\Phi_{xt}+\Phi_{x\theta_2}\right)-\sigma\left(\theta_1\Phi_{xt}+\Phi_{t\theta_1}\right)-\Pi\left(\cos{\Phi}\right)\\ &+\Pi^{xt}\left(\theta_1\theta_2\right)+\Pi^{t\theta_1}\left(\theta_2\right)-\Pi^{x\theta_2}\left(\theta_1\right)-\Pi^{\theta_1\theta_2}=0.
\end{split}
\label{symmie7}
\end{equation}
Note that proper respect to the ordering of odd terms is essential, e.g. $\Pi^{t\theta_1}$ is odd. We see that we only need to calculate the coefficients $\Pi^x$, $\Pi^t$, $\Pi^{\theta_1}$, $\Pi^{\theta_2}$, $\Pi^{xt}$, $\Pi^{t\theta_1}$, $\Pi^{x\theta_2}$ and $\Pi^{\theta_1\theta_2}$ in equation (\ref{symmie6}). They are found from the superspace version of the formulae for the 1st and 2nd prolongation of vector fields (see equation (\ref{prbos}) above)
\begin{equation}
 \Pi^{A}={\mathcal D}_{A} \Pi - \sum_{B} {\mathcal D}_{A} \zeta^B \Phi_B, \qquad 
\Pi^{AB}={\mathcal D}_{B} \Pi^A - \sum_{C} {\mathcal D}_{B} \zeta^C \Phi_{AC},
\label{symmie7A}
\end{equation}
where
\begin{equation}
 A,B,C\in\{x,t,\theta_1,\theta_2\}, \qquad \zeta^A=(\xi,\tau,\rho,\sigma).
\end{equation}
The derivation of these formulae is performed in the same way as in the bosonic case, working with infinitesimal transformations and keeping track of ordering properties. Explicitly, the coefficients are given as follows:
\begin{equation}\nonumber
\begin{split}
\Pi^x=&\Pi_x+\Pi_{\Phi}\Phi_x-\xi_x\Phi_x-\xi_{\Phi}(\Phi_x)^2-\tau_x\Phi_t-\tau_{\Phi}\Phi_x\Phi_t-\rho_x\Phi_{\theta_1}-\rho_{\Phi}\Phi_x\Phi_{\theta_1}-\sigma_x\Phi_{\theta_2}\\ &-\sigma_{\Phi}\Phi_x\Phi_{\theta_2},
\end{split}
\label{littlecoeff1}
\end{equation}

\begin{equation}\nonumber
\begin{split}
\Pi^t=&\Pi_t+\Pi_{\Phi}\Phi_t-\xi_t\Phi_x-\xi_{\Phi}\Phi_x\Phi_t-\tau_t\Phi_t-\tau_{\Phi}(\Phi_t)^2-\rho_t\Phi_{\theta_1}-\rho_{\Phi}\Phi_t\Phi_{\theta_1}-\sigma_t\Phi_{\theta_2}\\ &-\sigma_{\Phi}\Phi_t\Phi_{\theta_2},
\end{split}
\label{littlecoeff2}
\end{equation}

\begin{equation}\nonumber
\begin{split}
\Pi^{\theta_1}=&\Pi_{\theta_1}+\Pi_{\Phi}\Phi_{\theta_1}-\xi_{\theta_1}\Phi_x-\xi_{\Phi}\Phi_x\Phi_{\theta_1}-\tau_{\theta_1}\Phi_t-\tau_{\Phi}\Phi_t\Phi_{\theta_1}-\rho_{\theta_1}\Phi_{\theta_1}-\sigma_{\theta_1}\Phi_{\theta_2}\\ &+\sigma_{\Phi}\Phi_{\theta_1}\Phi_{\theta_2},
\end{split}
\label{littlecoeff3}
\end{equation}

\begin{equation}\nonumber
\begin{split}
\Pi^{\theta_2}=&\Pi_{\theta_2}+\Pi_{\Phi}\Phi_{\theta_2}-\xi_{\theta_2}\Phi_x-\xi_{\Phi}\Phi_x\Phi_{\theta_2}-\tau_{\theta_2}\Phi_t-\tau_{\Phi}\Phi_t\Phi_{\theta_2}-\rho_{\theta_2}\Phi_{\theta_1}-\rho_{\Phi}\Phi_{\theta_1}\Phi_{\theta_2}\\ &-\sigma_{\theta_2}\Phi_{\theta_2},
\end{split}
\label{littlecoeff4}
\end{equation}

\begin{equation}
\begin{split}
\Pi^{xt}=&\Pi_{xt}+\Pi_{x\Phi}\Phi_t+\Pi_{t\Phi}\Phi_x+\Pi_{\Phi\Phi}\Phi_x\Phi_t+\Pi_{\Phi}\Phi_{xt}-\xi_{xt}\Phi_x-\xi_{x\Phi}\Phi_x\Phi_t-\xi_x\Phi_{xt}\\ &-\xi_{t\Phi}(\Phi_x)^2-\xi_{\Phi\Phi}(\Phi_x)^2\Phi_t-2\xi_{\Phi}\Phi_x\Phi_{xt}-\xi_t\Phi_{xx}-\xi_{\Phi}\Phi_t\Phi_{xx}-\tau_{xt}\Phi_t-\tau_{t\Phi}\Phi_x\Phi_t\\ &-\tau_t\Phi_{xt}-\tau_{x\Phi}(\Phi_t)^2-\tau_{\Phi\Phi}(\Phi_t)^2\Phi_x-2\tau_{\Phi}\Phi_t\Phi_{xt}-\tau_x\Phi_{tt}-\tau_{\Phi}\Phi_x\Phi_{tt}-\rho_{xt}\Phi_{\theta_1}\\ &-\rho_{x\Phi}\Phi_t\Phi_{\theta_1}-\rho_{t\Phi}\Phi_x\Phi_{\theta_1}-\rho_x\Phi_{t\theta_1}-\rho_t\Phi_{x\theta_1}-\rho_{\Phi\Phi}\Phi_x\Phi_t\Phi_{\theta_1}-\rho_{\Phi}\Phi_{xt}\Phi_{\theta_1}\\ &-\rho_{\Phi}\Phi_{t\theta_1}\Phi_x-\rho_{\Phi}\Phi_{x\theta_1}\Phi_t-\sigma_{xt}\Phi_{\theta_2}-\sigma_{x\Phi}\Phi_t\Phi_{\theta_2}-\sigma_{t\Phi}\Phi_x\Phi_{\theta_2}-\sigma_x\Phi_{t\theta_2}-\sigma_t\Phi_{x\theta_2}\\ &-\sigma_{\Phi\Phi}\Phi_x\Phi_t\Phi_{\theta_2}-\sigma_{\Phi}\Phi_{xt}\Phi_{\theta_2}-\sigma_{\Phi}\Phi_{t\theta_2}\Phi_x-\sigma_{\Phi}\Phi_{x\theta_2}\Phi_t,\end{split}
\label{bigcoeff1}
\end{equation}

\begin{equation}\nonumber
\begin{split}
\Pi^{t\theta_1}=&\Pi_{t\theta_1}+\Pi_{t\Phi}\Phi_{\theta_1}+\Pi_{\theta_1\Phi}\Phi_t+\Pi_{\Phi\Phi}\Phi_t\Phi_{\theta_1}+\Pi_{\Phi}\Phi_{t\theta_1}-\xi_{t\theta_1}\Phi_x-\xi_{t\Phi}\Phi_x\Phi_{\theta_1}-\xi_t\Phi_{x\theta_1}\\ &-\xi_{\theta_1\Phi}\Phi_x\Phi_t-\xi_{\Phi\Phi}\Phi_x\Phi_t\Phi_{\theta_1}-\xi_{\Phi}\Phi_t\Phi_{x\theta_1}-\xi_{\Phi}\Phi_x\Phi_{t\theta_1}-\xi_{\theta_1}\Phi_{xt}-\xi_{\Phi}\Phi_{xt}\Phi_{\theta_1}\\ &-\tau_{t\theta_1}\Phi_t-\tau_{t\Phi}\Phi_t\Phi_{\theta_1}-\tau_t\Phi_{t\theta_1}-\tau_{\theta_1\Phi}(\Phi_t)^2-\tau_{\Phi\Phi}(\Phi_t)^2\Phi_{\theta_1}-2\tau_{\Phi}\Phi_t\Phi_{t\theta_1}\\ &-\tau_{\theta_1}\Phi_{tt}-\tau_{\Phi}\Phi_{tt}\Phi_{\theta_1}-\rho_{t\theta_1}\Phi_{\theta_1}-\rho_{\theta_1\Phi}\Phi_t\Phi_{\theta_1}-\rho_{\theta_1}\Phi_{t\theta_1}-\sigma_{t\theta_1}\Phi_{\theta_2}+\sigma_{t\Phi}\Phi_{\theta_1}\Phi_{\theta_2}\\ &-\sigma_t\Phi_{\theta_1\theta_2}-\sigma_{\theta_1\Phi}\Phi_t\Phi_{\theta_2}+\sigma_{\Phi\Phi}\Phi_t\Phi_{\theta_1}\Phi_{\theta_2}+\sigma_{\Phi}\Phi_{t\theta_1}\Phi_{\theta_2}-\sigma_{\Phi}\Phi_t\Phi_{\theta_1\theta_2}-\sigma_{\theta_1}\Phi_{t\theta_2}\\ &+\sigma_{\Phi}\Phi_{\theta_1}\Phi_{t\theta_2},
\end{split}
\label{bigcoeff2}
\end{equation}

\begin{equation}\nonumber
\begin{split}
\Pi^{x\theta_2}=&\Pi_{x\theta_2}+\Pi_{x\Phi}\Phi_{\theta_2}+\Pi_{\theta_2\Phi}\Phi_x+\Pi_{\Phi\Phi}\Phi_x\Phi_{\theta_2}+\Pi_{\Phi}\Phi_{x\theta_2}-\xi_{x\theta_2}\Phi_x-\xi_{x\Phi}\Phi_x\Phi_{\theta_2}\\ &-\xi_x\Phi_{x\theta_2}-\xi_{\theta_2\Phi}(\Phi_x)^2-\xi_{\Phi\Phi}(\Phi_x)^2\Phi_{\theta_2}-2\xi_{\Phi}\Phi_x\Phi_{x\theta_2}-\xi_{\theta_2}\Phi_{xx}-\xi_{\Phi}\Phi_{xx}\Phi_{\theta_2}\\ &-\tau_{x\theta_2}\Phi_t-\tau_{x\Phi}\Phi_t\Phi_{\theta_2}-\tau_x\Phi_{t\theta_2}-\tau_{\theta_2\Phi}\Phi_x\Phi_t-\tau_{\Phi\Phi}\Phi_x\Phi_t\Phi_{\theta_2}-\tau_{\Phi}\Phi_x\Phi_{t\theta_2}\\ &-\tau_{\Phi}\Phi_t\Phi_{x\theta_2}-\tau_{\theta_2}\Phi_{xt}-\tau_{\Phi}\Phi_{xt}\Phi_{\theta_2}-\rho_{x\theta_2}\Phi_{\theta_1}+\rho_{x\Phi}\Phi_{\theta_2}\Phi_{\theta_1}+\rho_x\Phi_{\theta_1\theta_2}\\ &-\rho_{\theta_2\Phi}\Phi_x\Phi_{\theta_1}+\rho_{\Phi\Phi}\Phi_x\Phi_{\theta_2}\Phi_{\theta_1}+\rho_{\Phi}\Phi_{x\theta_2}\Phi_{\theta_1}+\rho_{\Phi}\Phi_x\Phi_{\theta_1\theta_2}-\rho_{\theta_2}\Phi_{x\theta_1}\\ &+\rho_{\Phi}\Phi_{\theta_2}\Phi_{x\theta_1}-\sigma_{x\theta_2}\Phi_{\theta_2}-\sigma_{\theta_2\Phi}\Phi_x\Phi_{\theta_2}-\sigma_{\theta_2}\Phi_{x\theta_2},
\end{split}
\label{bigcoeff3}
\end{equation}

\begin{equation}\nonumber
\begin{split}
\Pi^{\theta_1\theta_2}=&\Pi_{\theta_1\theta_2}-\Pi_{\theta_1\Phi}\Phi_{\theta_2}+\Pi_{\theta_2\Phi}\Phi_{\theta_1}-\Pi_{\Phi\Phi}\Phi_{\theta_1}\Phi_{\theta_2}+\Pi_{\Phi}\Phi_{\theta_1\theta_2}-\xi_{\theta_1\theta_2}\Phi_x+\xi_{\theta_1\Phi}\Phi_x\Phi_{\theta_2}\\ &+\xi_{\theta_1}\Phi_{x\theta_2}-\xi_{\theta_2\Phi}\Phi_x\Phi_{\theta_1}+\xi_{\Phi\Phi}\Phi_x\Phi_{\theta_1}\Phi_{\theta_2}+\xi_{\Phi}\Phi_{\theta_1}\Phi_{x\theta_2}-\xi_{\Phi}\Phi_x\Phi_{\theta_1\theta_2}-\xi_{\theta_2}\Phi_{x\theta_1}\\ &-\xi_{\Phi}\Phi_{\theta_2}\Phi_{x\theta_1}-\tau_{\theta_1\theta_2}\Phi_t+\tau_{\theta_1\Phi}\Phi_t\Phi_{\theta_2}+\tau_{\theta_1}\Phi_{t\theta_2}-\tau_{\theta_2\Phi}\Phi_t\Phi_{\theta_1}+\tau_{\Phi\Phi}\Phi_t\Phi_{\theta_1}\Phi_{\theta_2}\\ &+\tau_{\Phi}\Phi_{\theta_1}\Phi_{t\theta_2}-\tau_{\Phi}\Phi_t\Phi_{\theta_1\theta_2}-\tau_{\theta_2}\Phi_{t\theta_1}-\tau_{\Phi}\Phi_{\theta_2}\Phi_{t\theta_1}-\rho_{\theta_1\theta_2}\Phi_{\theta_1}+\rho_{\theta_1\Phi}\Phi_{\theta_1}\Phi_{\theta_2}\\ &-\rho_{\theta_1}\Phi_{\theta_1\theta_2}-\sigma_{\theta_1\theta_2}\Phi_{\theta_2}+\sigma_{\theta_2\Phi}\Phi_{\theta_1}\Phi_{\theta_2}-\sigma_{\theta_2}\Phi_{\theta_1\theta_2}.
\end{split}
\label{bigcoeff4}
\end{equation}

Substituting the above formulae into equation (\ref{symmie7}) and replacing each term $\Phi_{\theta_1\theta_2}$ in the resulting expression by the terms $\theta_1\theta_2\Phi_{xt}-\theta_2\Phi_{t\theta_1}+\theta_1\Phi_{x\theta_2}-\sin{\Phi}$, we obtain a series of determining equations for the functions $\xi$, $\tau$, $\rho$, $\sigma$ and $\Pi$. The general solution of these determining equations is given by
\begin{equation}
\begin{split}
\xi(x,\theta_1)&=-2C_1x+C_2-\underline{D_1}\theta_1,\qquad
\tau(t,\theta_2)=2C_1t+C_3-\underline{D_2}\theta_2,\\
\rho(\theta_1)&=-C_1\theta_1+\underline{D_1},\qquad
\sigma(\theta_2)=C_1\theta_2+\underline{D_2},\qquad
\Pi=0,
\end{split}
\label{symmie8}
\end{equation}
where the parameters $C_1$, $C_2$, $C_3 \in \Lambda_{even}$, while $\underline{D_1}, \underline{D_2} \in \Lambda_{odd}$. Thus, the algebra of infinitesimal transformations is the even part of the Lie superalgebra $\mathfrak{S}$ over $\Lambda$ spanned by the following generators
\begin{equation}
\begin{split}
L=-2x\partial_x&+2t\partial_t-\theta_1\partial_{\theta_1}+\theta_2\partial_{\theta_2},\qquad P_x=\partial_x,\qquad P_t=\partial_t,\\ Q_x&=-\theta_1\partial_x+\partial_{\theta_1},\qquad Q_t=-\theta_2\partial_t+\partial_{\theta_2}.
\end{split}
\label{symmie9}
\end{equation}
The even generators $L$, $P_x$ and $P_t$ represent a dilation and translations in space and time respectively, while the odd generators $Q_x$ and $Q_t$ are simply the generators of supersymmetric transformations identified in Section \ref{sec2}. This means that we have recovered the full super--Poincar\'{e} algebra in $(1+1)$ dimensions which was expected. The commutation (and anticommutation in the case of two odd generators) relations of the Lie super algebra $\mathfrak{S}$ generated by the vector fields (\ref{symmie9}) are given in Table 3.

\begin{table}[htbp]
  \begin{center}
\caption{Supercommutation table for the Lie superalgebra $\mathfrak{S}$ spanned by the
  vector fields (\ref{symmie9})}
\vspace{5mm}
\setlength{\extrarowheight}{4pt}
\begin{tabular}{|c||c|c|c|c|c|}\hline
& $\mathbf{L}$ & $\mathbf{P_x}$ & $\mathbf{P_t}$ & $\mathbf{Q_x}$ & $\mathbf{Q_t}$\\[0.5ex]\hline\hline
$\mathbf{L}$ & $0$ & $2P_x$ & $-2P_t$ & $Q_x$ & $-Q_t$ \\\hline
$\mathbf{P_x}$ & $-2P_x$ & $0$ & $0$ & $0$ & $0$ \\\hline
$\mathbf{P_t}$ & $2P_t$ & $0$ & $0$ & $0$ & $0$ \\\hline
$\mathbf{Q_x}$ & $-Q_x$ & $0$ & $0$ & $-2P_x$ & $0$ \\\hline
$\mathbf{Q_t}$ & $Q_t$ & $0$ & $0$ & $0$ & $-2P_t$ \\\hline
\end{tabular}
  \end{center}
\end{table}

\section{One--dimensional subalgebras of the symmetry algebra of the SSG equation}\label{1dsaSSG}

In this section, we classify the one--dimensional subalgebras of the Lie algebra of infinitesimal transformations $\mathfrak{S}_{even}$ into conjugacy classes under the action of the super Lie group ${\rm exp} (\mathfrak{S}_{even})$ generated by $\mathfrak{S}_{even}$. Such a classification is of importance for us because conjugate subgroups necessarily lead to invariant solutions equivalent in the sense that they can be transformed by a suitable symmetry from one to the other -- therefore, there is no need to compute reductions with respect to algebras which are conjugate to each other. On the other hand, for our purposes it is not of particular importance to establish exactly one representative of each class, as long as the procedure of reduction has the same form for all the representatives, differing by a choice of parameters only.

We recall why $\mathfrak{S}_{even}$ is the algebra we are interested in. It would be inconsistent to consider the $\mathbb{R}$--span of the generators (\ref{symmie9}), because we multiply the odd generators $Q_x$ and $Q_t$ by the odd parameters $\underline{\eta_1}$ and $\underline{\eta_2}$ respectively in equation (\ref{b2}). Therefore, one is naturally led to consideration of $\mathfrak{S}_{even}$ which is a supermanifold in the sense presented in Section \ref{sec2}. It means that $\mathfrak{S}_{even}$ contains sums of any even combination of $P_x,P_t,L$ (i.e. multiplied by even parameters in $\Lambda_{even}$, including real numbers), and odd combination  of $Q_x$ and $Q_t$ (i.e. multiplied by odd parameters in $\Lambda_{odd}$). At the same time, $\mathfrak{S}_{even}$ is a $\Lambda_{even}$ Lie module.

This leads to the following complication:
\begin{itemize}
 \item for a given $X \in \mathfrak{S}$ the subalgebras $\mathfrak{X},\mathfrak{X}'$  spanned by $X$ and by $X'= a X,\, a \in \Lambda_{even} {\backslash} \R $ are in general not isomorphic, $\mathfrak{X}' \subset \mathfrak{X}$.
\end{itemize}
It seems that the subalgebras obtained from other ones through multiplication by nilpotent elements of $\Lambda_{even}$ do not give us anything new for the purpose of symmetry reduction -- they may allow a bit more freedom in the choice of invariants but we then encounter the problem of non--standard invariants which we will discuss at the end of Section \ref{isSSG}. 

Similarly, it does not appear to be particularly useful to consider a subalgebra of the form e.g. $\{P_x+ {\underline\eta_1} {\underline\eta_2} P_t\}$ (although the reduction for this case can easily be reconstructed by substituting $\varepsilon={\underline\eta_1} {\underline\eta_2}$ in the subalgebra $\mathfrak{S}_4$ and the corresponding formulae below.)

Therefore,  we will assume throughout the computation of the non--isomorphic one--dimensional subalgebras that the nonzero even parameters are invertible, i.e. behave essentially like ordinary real numbers.

The Lie algebra $\mathfrak{S}_{even}$ can be decomposed into the semi-direct sum
\begin{equation}
\mathfrak{S}=\{L\}\sdir\{P_x,P_t,Q_x,Q_t\}.
\label{symmie10}
\end{equation}
In order to classify this Lie superalgebra, we make use of the techniques for semi--direct sums of algebras described in \cite{Winternitz} (Section 4.4) and generalize them to superalgebras involving both even and odd generators. Here, we identify the components $F$ and $N$ of the semi--direct sum as
\begin{displaymath}
F=\{L\},\qquad N=\{P_x,P_t,Q_x,Q_t\}.
\end{displaymath}
The trivial subalgebras of $F$ are simply $F_1=\{0\}$ and $F_2=\{L\}$. We begin by considering the splitting one-dimensional subalgebras.

For $F_1=\{0\}$, all one-dimensional subspaces of the form
\begin{equation}
\{\alpha P_x+\beta P_t+\underline{\mu}Q_x+\underline{\nu}Q_t\},\mbox{ }\alpha,\beta \in \Lambda_{even},\ \underline{\mu},\underline{\nu} \in \Lambda_{odd}
\label{symmie11}
\end{equation}
are invariant subalgebras, i.e. subalgebras of $N$ invariant under the action of $F_1$. 

Under the action of the one--parameter group generated by the  generator
\begin{equation}
Y=kL+mP_x+nP_t+\underline{\eta}Q_x+\underline{\lambda}Q_t,
\label{symmie12}
\end{equation}
where $k$, $m$, $n\in \Lambda_{even}$ and $\underline{\eta}$, $\underline{\lambda}\in \Lambda_{odd}$, the one--dimensional subalgebra (\ref{symmie11}) transforms under the Baker--Campbell--Hausdorff formula
\begin{equation}
X\longrightarrow {\rm Ad}_{exp(Y)} X = X+[Y,X]+\frac{1}{2!}\left[Y,[Y,X]\right]+\frac{1}{ 3!}\left[Y,\left[Y,[Y,X]\right]\right]+\ldots
\label{symmie13}
\end{equation}
to
\begin{equation}
\left(e^{2k}\alpha+{2}\underline{\eta}\underline{\mu}e^k \, \frac{\left(e^{k}-1\right)}{k} \right) P_x+\left(e^{-2k}\beta+{2}\underline{\lambda}\underline{\nu}
e^{-2k} \, \frac{\left(e^{k}-1\right)}{k}\right)P_t+e^k\underline{\mu}Q_x+e^{-k}\underline{\nu}Q_t.
\label{symmie14}
\end{equation}
If $k$ is bodiless (see equation (\ref{bodysoul})) then we interpret $\frac{e^{k}-1}{k}$ as its well--defined limit
$ \frac{e^{k}-1}{k} = \sum_{j=0}^{\infty} \frac{1}{(j+1)!} k^j.$

We note that the action (\ref{symmie13}) with $Y=kL$ on the even generators $\alpha P_x+\beta P_t$ together with an overall rescaling of the subalgebra generator can be always used to bring one of the coefficients $\alpha,\beta$ to 1 under the assumption that at least one of them was invertible supernumber. The other can be scaled either to $\pm 1$ or bodiless even supernumber. Note that here the assumption of finite number of Grassmann generators of $\Lambda$ is essential -- it guarantees a cut off in the sum $ \ln(1+\gamma)=\sum_{j\geq 1} \frac{(-1)^{j-1}}{j} \gamma^j$ for bodiless $\gamma\in \Lambda_{even}$ so that one doesn't have to worry about its convergence; i.e. $k\in\Lambda_{even}$ such that $e^k=1+\gamma$ exists for every even bodiless $\gamma$. 

Once the coefficients $P_x,P_t$ are brought to simple form, one uses any remaining freedom to simplify the coefficients of $Q_x,Q_t$. As is seen from equation (\ref{symmie14}) not much can be accomplished -- only rescaling by $\exp(k), \, k \in \Lambda_{even}$ (and an overall rescaling if both $\alpha=\beta=0$) may still be available.

Considering first the subalgebras containing only the generators $P_x,P_t$, we obtain essentially the same subalgebras as for the system in component form described in Section \ref{LsssGcf}.\\\\
{\bf (i)} If $\beta=0$, $\underline{\mu}=0$, $\underline{\nu}=0$, we have the subalgebra $\{P_x\}$ which  is not conjugate to any other subalgebra.\\\\
{\bf (ii)} If $\alpha=0$, $\underline{\mu}=0$, $\underline{\nu}=0$, we have the subalgebra $\{P_t\}$.\\\\
{\bf (iii)} The subalgebra $\{\alpha P_x+\beta P_t\}$, where $a,b\in\Lambda_{even}$ such that $a^{-1}$ exists can be brought to the form  
\begin{equation}\label{PeP}
\{P_x+\varepsilon P_t\}, 
\end{equation}
 where $\varepsilon=\pm 1$ or $\varepsilon$ is bodiless. If $a$ is bodiless then we get similarly $$\{P_t+\omega P_x\},$$
where $\omega$ is bodiless. As mentioned at the beginning of this section, we shall consider only subalgebra (\ref{PeP}) with $\epsilon=\pm 1$ in what follows.

Next we complement the generators $P_x,P_t$ by $Q_x,Q_t$. This leads to the following non--conjugate subalgebra types\\\\
{\bf (iv)}  $\{\underline{\mu}Q_x\}$,
\\\\
{\bf (v)} $\{\underline{\nu}Q_t\}$,
\\\\
{\bf (vi)} $\{P_x+\underline{\mu}Q_x\}$ where algebras with $\underline{\mu}$ and $e^{k} \underline{\mu}, \, k \in \Lambda_{even}$ are isomorphic,
\\\\
{\bf (vii)} $\{P_t+\underline{\mu}Q_x\}$ where algebras with $\underline{\mu}$ and $e^{k} \underline{\mu}, \, k \in \Lambda_{even}$ are isomorphic,\\\\
{\bf (viii)} $\{P_x+\underline{\nu}Q_t\}$ where algebras with $\underline{\nu}$ and $e^{k} \underline{\nu}, \, k \in \Lambda_{even}$ are isomorphic,\\\\
{\bf (ix)} $\{P_t+\underline{\nu}Q_t\}$ where algebras with $\underline{\nu}$ and $e^{k} \underline{\nu}, \, k \in \Lambda_{even}$ are isomorphic,\\\\
{\bf (x)}  $\{P_x+\varepsilon P_t+\underline{\mu}Q_x\}$, \\\\
{\bf (xi)} $\{P_x+\varepsilon P_t+\underline{\nu}Q_t\}$,\\\\
{\bf (xii)} $\{\underline{\mu}Q_x+\underline{\nu}Q_t\}$ where both $\underline{\mu},\underline{\nu}$ can be simultaneously rescaled by $a\in\Lambda_{even}$ and then one of them by $e^k,\, k\in\Lambda_{even}$, \\\\
{\bf (xiii)} $\{P_x+\underline{\mu}Q_x+\underline{\nu}Q_t\}$ where algebras defined by  $(\underline{\mu},\underline{\nu})$ and $(e^{k}\underline{\mu},e^{3k}\underline{\nu}), \, k \in \Lambda_{even}$ are isomorphic,
\\\\
{\bf (xiv)} $\{P_t+\underline{\mu}Q_x+\underline{\nu}Q_t\}$  where algebras defined by  $(\underline{\mu},\underline{\nu})$ and $(e^{3k}\underline{\mu},e^{k}\underline{\nu}), \, k \in \Lambda_{even}$ are isomorphic, 
\\\\
{\bf (xv)} $\{P_x+\varepsilon P_t+\underline{\mu}Q_x+\underline{\nu}Q_t\}$.

For $F_2=\{L\}$, the only splitting one-dimensional subalgebra is $\{L\}$ itself.

Next, we then look for non--splitting subalgebras of $\mathfrak{S}$ of the form
\begin{equation}
V=\{L+\alpha P_x+\beta P_t+\underline{\mu}Q_x+\underline{\nu}Q_t\},
\label{symmie32}
\end{equation}
but an easy calculation using Baker--Campbell--Hausdorff formula (\ref{symmie13}) shows that all such algebras are conjugate to $\{L\}$. Thus, there are no separate conjugacy classes of non--splitting one--dimensional subalgebras of $\mathfrak{S}$.

Therefore, the one--dimensional subalgebra classification (under the restrictions mentioned at the beginning of this section) is
\begin{equation}
\begin{split}
&\mathfrak{S}_1=\{L\},\qquad \mathfrak{S}_2=\{P_x\},\qquad \mathfrak{S}_3=\{P_t\},\qquad \mathfrak{S}_4=\{P_x+\varepsilon P_t\},\qquad \mathfrak{S}_5=\{\underline{\mu}Q_x\},\\ &\mathfrak{S}_6=\{P_x+\underline{\mu}Q_x\},\qquad \mathfrak{S}_7=\{P_t+\underline{\mu}Q_x\},\qquad \mathfrak{S}_8=\{P_x+\varepsilon P_t+\underline{\mu}Q_x\},\\ &\mathfrak{S}_9=\{\underline{\nu}Q_t\},\qquad \mathfrak{S}_{10}=\{P_x+\underline{\nu}Q_t\},\qquad \mathfrak{S}_{11}=\{P_t+\underline{\nu}Q_t\},\\ &\mathfrak{S}_{12}=\{P_x+\varepsilon P_t+\underline{\nu}Q_t\},\qquad \mathfrak{S}_{13}=\{\underline{\mu}Q_x+\underline{\nu}Q_t\},\qquad \mathfrak{S}_{14}=\{P_x+\underline{\mu}Q_x+\underline{\nu}Q_t\},\\ &\mathfrak{S}_{15}=\{P_t+\underline{\mu}Q_x+\underline{\nu}Q_t\},\qquad \mathfrak{S}_{16}=\{P_x+\varepsilon P_t+\underline{\mu}Q_x+\underline{\nu}Q_t\}.
\end{split}
\label{symmie33}
\end{equation}
Any parameter, if present, is assumed to be nonvanishing. Underlined parameters belong to $\Lambda_{odd}$, $\varepsilon=\pm 1$ (although also $\varepsilon\in\Lambda_{even}$ bodiless can in principle be considered.)

This classification will allow us to use the SRM in order to determine invariant solutions of the SSG equation (\ref{symmie2}).

\section{Invariant solutions of the supersymmetric sine--Gordon equation}\label{isSSG}

We now proceed to apply a modified version of the SRM to the SSG equation (\ref{symmie2}) in order to obtain invariant solutions of the model. Considering in turn each of the one-dimensional subalgebras described in Section \ref{1dsaSSG}, we begin by constructing, where possible, a set of four independent invariants of the specific subalgebra. In each case, the even invariant is labelled by $\sigma$ and the odd invariant(s) by $\tau$ (or $\tau_1$, $\tau_2$). For the subalgebras $\mathfrak{S}_5$, $\mathfrak{S}_9$, $\mathfrak{S}_{13}$, $\mathfrak{S}_{14}$, $\mathfrak{S}_{15}$ and $\mathfrak{S}_{16}$, the structure of the invariants is non--standard and will be discussed at the end of this section.

The bosonic superfield $\Phi$ is expanded in terms of its various odd invariants. The dependence of $\Phi$ on each odd variable $\tau_i$ must be at most linear (as $(\tau_i)^2=0$). Substituting this decomposition into the SSG equation (\ref{symmie2}), we obtain a reduced partial differential equation  for the superfield $\Phi$ which in turn leads us to a system of differential constraints between its component even and odd functions. For instance, if the invariants are given by $\sigma$, $\tau_1$, $\tau_2$, $\Phi$, the superfield $\Phi$ can be decomposed into the form
\begin{equation}
\Phi={\mathcal A}\left(\sigma,\tau_1,\tau_2\right)=\alpha(\sigma)+\tau_1\eta(\sigma)+\tau_2\lambda(\sigma)+\tau_1\tau_2\beta(\sigma),
\label{ginv1}
\end{equation}
where $\alpha$ and $\beta$ are  even--valued functions of $\sigma$ while $\eta$ and $\lambda$ are  odd--valued functions of $\sigma$. Substitution into the SSG equation (\ref{symmie2}) allows us to determine the differential constraints between the functions $\alpha$, $\beta$, $\eta$ and $\lambda$. In general, the reduced supersymmetric equation will contain the term $\sin{\mathcal A}$ which can be expanded in the form
\begin{equation}
\sin{\mathcal A}=\sin{\alpha}+\tau_1\eta\cos{\alpha}+\tau_2\lambda\cos{\alpha}+\tau_1\tau_2\left(\beta\cos{\alpha}+\eta\lambda\sin{\alpha}\right),
\label{ginv2}
\end{equation}
as identified from the series
\begin{equation}
\sin{\mathcal A}={\mathcal A}-\frac{1}{3!}{\mathcal A}^3+\frac{1}{5!}{\mathcal A}^5 - \ldots
\label{ginv3}
\end{equation}
The results are summarized in Tables 4 and 5. In Table 4, we list the one-dimensional subalgebras and their respective invariants and superfields. In Table 5, we present the systems of differential constraints resulting from each symmetry reduction and assumed form of the superfield. In what follows, we deal separately with each case described above by performing an analysis of the various solutions of the obtained differential constraints. The resulting expressions are then substituted into the superfield formula for $\Phi$, from which we obtain group-invariant solutions.

\begin{table}[htbp]
  \begin{center}
\caption{Invariants and change of variables for subalgebras of the Lie superalgebra $\mathfrak{S}$ spanned by the  vector fields (\ref{symmie9})}
\vspace{3mm}
\setlength{\extrarowheight}{4pt}
\begin{tabular}{|c|c|c|}\hline
Subalgebra & Invariants & Superfield\\[0.5ex]\hline\hline
$\mathfrak{S}_1=\{L\}$ & $\sigma=xt$, $\tau_1=t^{1/2}\theta_1$, & $\Phi={\mathcal A}\left(\sigma,\tau_1,\tau_2\right)=\alpha(\sigma)+\tau_1\mu(\sigma)+\tau_2\nu(\sigma)+\tau_1\tau_2\beta(\sigma)$ \\
 & $\tau_2=t^{-1/2}\theta_2$, $\Phi$ &\\\hline
$\mathfrak{S}_2=\{P_x\}$ & $t$, $\theta_1$, $\theta_2$, $\Phi$  & $\Phi={\mathcal A}\left(t,\theta_1,\theta_2\right)=\alpha(t)+\theta_1\mu(t)+\theta_2\nu(t)+\theta_1\theta_2\beta(t)$ \\\hline
$\mathfrak{S}_3=\{P_t\}$ & $x$, $\theta_1$, $\theta_2$, $\Phi$  & $\Phi={\mathcal A}\left(x,\theta_1,\theta_2\right)=\alpha(x)+\theta_1\mu(x)+\theta_2\nu(x)+\theta_1\theta_2\beta(x)$ \\\hline
$\mathfrak{S}_4=\{P_x+\varepsilon P_t\}$ & $\sigma=x-\varepsilon t$, $\theta_1$, $\theta_2$, $\Phi$  & $\Phi={\mathcal A}\left(\sigma,\theta_1,\theta_2\right)=\alpha(\sigma)+\theta_1\mu(\sigma)+\theta_2\nu(\sigma)+\theta_1\theta_2\beta(\sigma)$ \\\hline
$\mathfrak{S}_6=\{P_x+\underline{\mu}Q_x\}$ & $t$, $\tau=\theta_1-\underline{\mu}x$, $\theta_2$, $\Phi$  & $\Phi={\mathcal A}\left(t,\tau,\theta_2\right)=\alpha(t)+\tau\eta(t)+\theta_2\lambda(t)+\tau\theta_2\beta(t)$ \\\hline
$\mathfrak{S}_7=\{P_t+\underline{\mu}Q_x\}$ & $\sigma=x+\underline{\mu}\theta_1t$, & $\Phi={\mathcal A}\left(\sigma,\tau,\theta_2\right)=\alpha(\sigma)+\tau\eta(\sigma)+\theta_2\lambda(\sigma)+\tau\theta_2\beta(\sigma)$ \\
 & $\tau=\theta_1-\underline{\mu}t$, $\theta_2$, $\Phi$ & \\\hline
$\mathfrak{S}_8=\{P_x+\varepsilon P_t+\underline{\mu}Q_x\}$ & $\sigma=\varepsilon x-t+\underline{\mu}t\theta_1$, & $\Phi={\mathcal A}\left(\sigma,\tau,\theta_2\right)=\alpha(\sigma)+\tau\eta(\sigma)+\theta_2\lambda(\sigma)+\tau\theta_2\beta(\sigma)$ \\
& $\tau=\theta_1-\varepsilon\underline{\mu}t$, $\theta_2$, $\Phi$ & \\\hline
$\mathfrak{S}_{10}=\{P_x+\underline{\nu}Q_t\}$ & $\sigma=t+\underline{\nu}\theta_2x$, & $\Phi={\mathcal A}\left(\sigma,\tau,\theta_1\right)=\alpha(\sigma)+\tau\eta(\sigma)+\theta_1\lambda(\sigma)+\tau\theta_1\beta(\sigma)$ \\
 & $\tau=\theta_2-\underline{\nu}x$, $\theta_1$, $\Phi$ & \\\hline
$\mathfrak{S}_{11}=\{P_t+\underline{\nu}Q_t\}$ & $x$, $\theta_1$, $\tau=\theta_2-\underline{\nu}t$, $\Phi$  & $\Phi={\mathcal A}\left(x,\tau,\theta_1\right)=\alpha(x)+\tau\eta(x)+\theta_1\lambda(x)+\tau\theta_1\beta(x)$ \\\hline
$\mathfrak{S}_{12}=\{P_x+\varepsilon P_t+\underline{\nu}Q_t\}$ & $\sigma=t-\varepsilon x+\underline{\nu}x\theta_2$, & $\Phi={\mathcal A}\left(\sigma,\tau,\theta_1\right)=\alpha(\sigma)+\tau\eta(\sigma)+\theta_1\lambda(\sigma)+\tau\theta_1\beta(\sigma)$ \\
 & $\tau=\theta_2-\underline{\nu}x$, $\theta_1$, $\Phi$ &\\\hline
\end{tabular}
  \end{center}
\end{table}

\begin{table}[htbp]
  \begin{center}
\caption{Reduced equations obtained for subalgebras of the Lie superalgebra $\mathfrak{S}$ spanned by the
  vector fields (\ref{symmie9})}\label{sre}
\vspace{3mm}
\setlength{\extrarowheight}{4pt}
\begin{tabular}{|c|c|}\hline
Subalgebra & Reduced equations \\[0.5ex]\hline\hline
$\mathfrak{S}_1=\{L\}$ & $\beta+\sin{\alpha}=0$,\qquad $\nu_{\sigma}-\mu\cos{\alpha}=0$,\\ & $\sigma\mu_{\sigma}+\frac{1}{2}\mu+\nu\cos{\alpha}=0$,\qquad $\alpha_{\sigma}+\sigma\alpha_{\sigma\sigma}-\beta\cos{\alpha}-\mu\nu\sin{\alpha}=0$ \\\hline
$\mathfrak{S}_2=\{P_x\}$ & $\beta+\sin{\alpha}=0$,\qquad $\mu\cos{\alpha}=0$,\\ & $\mu_t+\nu\cos{\alpha}=0$,\qquad $\beta\cos{\alpha}+\mu\nu\sin{\alpha}=0$ \\\hline
$\mathfrak{S}_3=\{P_t\}$ & $\beta+\sin{\alpha}=0$,\qquad $\nu_x-\mu\cos{\alpha}=0$,\\ & $\nu\cos{\alpha}=0$,\qquad $\beta\cos{\alpha}+\mu\nu\sin{\alpha}=0$ \\\hline
$\mathfrak{S}_4=\{P_x+\varepsilon P_t\}$ & $\beta+\sin{\alpha}=0$,\qquad $\nu_{\sigma}-\mu\cos{\alpha}=0$,\\ & $\varepsilon\mu_{\sigma}-\nu\cos{\alpha}=0$,\qquad $\varepsilon\alpha_{\sigma\sigma}+\beta\cos{\alpha}+\mu\nu\sin{\alpha}=0$ \\\hline
$\mathfrak{S}_6=\{P_x+\underline{\mu}Q_x\}$ & $\beta+\sin{\alpha}=0$,\qquad $\underline{\mu}\beta-\eta\cos{\alpha}=0$,\\ & $\eta_t+\lambda\cos{\alpha}=0$,\qquad $\underline{\mu}\eta_t+\beta\cos{\alpha}+\eta\lambda\sin{\alpha}=0$ \\\hline
$\mathfrak{S}_7=\{P_t+\underline{\mu}Q_x\}$ & $\beta+\sin{\alpha}=0$,\qquad $\lambda_{\sigma}-\eta\cos{\alpha}=0$,\\ & $\underline{\mu}\alpha_{\sigma}-\lambda\cos{\alpha}=0$,\qquad $\underline{\mu}\eta_{\sigma}+\beta\cos{\alpha}+\eta\lambda\sin{\alpha}=0$ \\\hline
$\mathfrak{S}_8=\{P_x+\varepsilon P_t+\underline{\mu}Q_x\}$ & $\beta+\sin{\alpha}=0$,\qquad $\varepsilon\lambda_{\sigma}-\eta\cos{\alpha}=0$,\\ & $\eta_{\sigma}+\underline{\mu}\alpha_{\sigma}-\lambda\cos{\alpha}=0$,\qquad $\varepsilon\alpha_{\sigma\sigma}+\underline{\mu}\eta_{\sigma}+\beta\cos{\alpha}+\eta\lambda\sin{\alpha}=0$ \\\hline
$\mathfrak{S}_{10}=\{P_x+\underline{\nu}Q_t\}$ & $\beta-\sin{\alpha}=0$,\qquad $\lambda_{\sigma}+\eta\cos{\alpha}=0$,\\ & $\underline{\nu}\alpha_{\sigma}+\lambda\cos{\alpha}=0$,\qquad $\underline{\nu}\eta_{\sigma}-\beta\cos{\alpha}-\eta\lambda\sin{\alpha}=0$ \\\hline
$\mathfrak{S}_{11}=\{P_t+\underline{\nu}Q_t\}$ & $\beta-\sin{\alpha}=0$,\qquad $\underline{\nu}\beta+\eta\cos{\alpha}=0$,\\ & $\eta_x-\lambda\cos{\alpha}=0$,\qquad $\underline{\nu}\eta_x-\beta\cos{\alpha}-\eta\lambda\sin{\alpha}=0$ \\\hline
$\mathfrak{S}_{12}=\{P_x+\varepsilon P_t+\underline{\nu}Q_t\}$ & $\beta-\sin{\alpha}=0$,\qquad $\lambda_{\sigma}+\eta\cos{\alpha}=0$,\\ & $\underline{\nu}\alpha_{\sigma}+\varepsilon\eta_{\sigma}+\lambda\cos{\alpha}=0$,\qquad $\varepsilon\alpha_{\sigma\sigma}+\underline{\nu}\eta_{\sigma}-\beta\cos{\alpha}-\eta\lambda\sin{\alpha}=0$ \\\hline
\end{tabular}
  \end{center}
\end{table}

Subalgebra $\mathfrak{S}_1=\{L\}$ leads to the reduction
\begin{equation}
\Phi(x,t,\theta_1,\theta_2)=\alpha(\sigma)+t^{1/2}\theta_1\mu(\sigma)+t^{-1/2}\theta_2\nu(\sigma)+\theta_1\theta_2\beta(\sigma),
\label{ginv4}
\end{equation}
where $\sigma=xt$ and the functions $\alpha$, $\beta$, $\mu$, $\nu$ satisfy
\begin{equation}\label{d7}
\begin{split}
&\sigma\alpha_{\sigma\sigma}+\alpha_{\sigma}+\frac{1}{2}\sin{(2\alpha)}-C_0\sigma^{-1/2}\sin{\alpha} =  0,\\
&\nu_{\sigma\sigma}+\tan{\alpha}\,\alpha_{\sigma} \,\nu_{\sigma}+\frac{1}{2\sigma}\nu_{\sigma}+\frac{1}{ \sigma}\cos^2{\alpha}\,\nu=0,\\
& \mu-\frac{1}{\cos{\alpha}}\nu_{\sigma}=0,\\
&\beta+\sin{\alpha}=0,\\
&(\sigma^{1/2} \mu\nu)_\sigma=0,
\end{split}
\end{equation}
where $C_0$ denotes the nilpotent even constant equal to $\sigma^{1/2} \mu\nu.$
These equations are equivalent to the ones listed in Table \ref{sre} above but written in a form more convenient for further simplification.
This reduction is also equivalent to that found for the SSG equation in the component form in Table \ref{tb2} in Section \ref{LsssGcf}. 

We find it convenient to consider the first equation in (\ref{d7}), namely
$$ \sigma\alpha_{\sigma\sigma}+\alpha_{\sigma}+\frac{1}{2}\sin{(2\alpha)}-C_0\sigma^{-1/2}\sin{\alpha} =  0 $$
 as a complex ordinary differential equation. Then under the transformation
\begin{equation}
\alpha=i\ln{y},
\label{d13}
\end{equation}
it becomes
\begin{equation}
y_{\sigma\sigma}={1\over y}\left(y_{\sigma}\right)^2-{1\over \sigma}y_{\sigma}+{1\over 4\sigma}\left(y^{-1}-y^3\right)-{C_0\over 2\sigma^{3/2}}\left(1-y^2\right).
\label{d14}
\end{equation}
In the case where $C_0=0$, we can re--scale the independent variable $\sigma$ to $z=\pm 2i\sigma$ and we obtain the following form of equation (\ref{d14}) 
\begin{equation}
y_{zz}={1\over y}\left(y_z\right)^2-{1\over z}y_z\pm{i\over 8z}\left(y^3-{1\over y}\right).
\label{d15}
\end{equation}
The solution of the reduced system (\ref{d7}) can be expressed in terms of $y$ through the transformation (\ref{d13}). Under the assumption that $C_0=0$ the odd--valued functions $\mu$ and $\nu$ have to satisfy the following differential equations
\begin{equation}
\begin{split}
&\nu_{\sigma\sigma}=-\left({1\over 2\sigma}+{1-y^2\over y(1+y^2)}y_{\sigma}\right)\nu_{\sigma}-{1\over 4\sigma}\left(y+{1\over y}\right)^2\nu,\qquad \mu={2y\over 1+y^2}\nu_{\sigma},
\end{split}
\label{d16}
\end{equation}
together with the constraint $\mu\nu=0$.

On the other hand, taking $\alpha=0$ in equation (\ref{d7}), we obtain the following particular solution of the  SSG equation
\begin{equation}
\begin{split}
\Phi(x,t,\theta_1,\theta_2)=\left[ {\underline{D_1}\over \sqrt{x}}\cos{\left(2\sqrt{xt}\right)}-{\underline{D_2}\over \sqrt{x}}\sin{\left(2\sqrt{xt}\right)} \right] \theta_1 + \left[{\underline{D_1}\over \sqrt{t}}\sin{\left(2\sqrt{xt}\right)}+{\underline{D_2}\over \sqrt{t}}\cos{\left(2\sqrt{xt}\right)}\right] \theta_2
\end{split}
\label{d18}
\end{equation}
representing a nonsingular periodic solution with damping factor $t^{-1/2}$ (where $t\neq 0$).

For subalgebra $\mathfrak{S}_2=\{P_x\}$ the reduced equations in Table \ref{sre} are equivalent to the corresponding ones obtained in component form, i.e. listed in Table \ref{tb2} in Section \ref{LsssGcf}. The only nonvanishing solutions which we obtain are
\begin{equation}
\Phi(x,t,\theta_1,\theta_2)=k\pi,
\label{gian1}
\end{equation}
where $k\in\mathbb{Z}$ and
\begin{equation}
\Phi(x,t,\theta_1,\theta_2)=(k+\frac{1}{2})\pi+\theta_1\underline{\mu_0}+\theta_2\underline{\mu_0}\varphi(t)+(-1)^{k+1}\theta_1\theta_2,
\label{gian1A}
\end{equation}
where $k\in\mathbb{Z}$, $\underline{\mu_0}$ is an odd supernumber and $\varphi$ is an arbitrary even--valued function of $t$. 

Similarly, in the case of subalgebra $\mathfrak{S}_3=\{P_t\}$  the reduced equations in Table \ref{sre} are equivalent to the corresponding ones obtained in component form in Table \ref{tb2}. The only non--zero solutions are
\begin{equation}
\Phi(x,t,\theta_1,\theta_2)=k\pi,
\label{gian1B}
\end{equation}
where $k\in\mathbb{Z}$ and
\begin{equation}
\Phi(x,t,\theta_1,\theta_2)=(k+\frac{1}{2})\pi+\theta_1\underline{\nu_0}\varphi(x)+\theta_2\underline{\nu_0}+(-1)^{k+1}\theta_1\theta_2,
\label{gian1C}
\end{equation}
where $k\in\mathbb{Z}$, $\underline{\nu_0}$ is an odd supernumber and $\varphi$ is an arbitrary even--valued function of $x$. 

Subalgebra $\mathfrak{S}_4=\{P_x+\varepsilon P_t\}$ leads to the last reduction which was obtained also in the component form in Table \ref{tb2} in Section \ref{LsssGcf}. The reduced equations for the superfield
\begin{equation}
\Phi(x,t,\theta_1,\theta_2)=\alpha(\sigma)+\theta_1\mu(\sigma)+\theta_2\nu(\sigma)+\theta_1\theta_2\beta(\sigma),
\label{ginv4a}
\end{equation}
where $\sigma=x-\varepsilon t$ are equivalent to the following set of equations for functions $\alpha,\nu,\mu,\beta$ 
\begin{equation}\label{res4}
\begin{split}
&\varepsilon\alpha_{\sigma\sigma}-\frac{1}{2}\sin{(2\alpha)}+K_0\sin{\alpha}=0,\\
&\nu_{\sigma\sigma}+\tan{\sigma}\nu_{\sigma}\alpha_{\sigma}-\varepsilon\cos^2{\alpha}\nu=0,\\
&\mu-\frac{1}{\cos{\alpha}}\nu_{\sigma}=0,\\
&\beta+\sin{\alpha}=0, \\
& (\mu\nu)_\sigma=0
\end{split}
\end{equation}
where we denoted the nilpotent constant $\mu\nu$ by $K_0$. The resulting solutions are traveling wave solutions in both the even and odd fields. We recall that the equation for $\alpha$, namely
\begin{equation}\label{rebp}
 \varepsilon\alpha_{\sigma\sigma}-\frac{1}{2}\sin{(2\alpha)}+K_0\sin{\alpha}=0 
\end{equation} 
appears in the reduction of the double sine-Gordon equation in $(2+1)$ dimensions \cite{Bullough} but with real $K_0$.
 
Considering the different values of $\epsilon$ separately, in the case $\epsilon=1$ we make the substitution $\alpha=-i\ln{v}$ into equation (\ref{rebp}) followed by an integration which leads to the equation
\begin{equation}
(v_{\sigma})^2-\frac{1}{4}\left(v^4-4 K_0 v^3+8 K_1 v^2- 4 K_0 v+1\right)=0,
\label{d2}
\end{equation}
solved by an elliptic integral in terms of a P--Weierstrass function. When $K_0=0$, equation (\ref{rebp}) reduces to the reduced sine--Gordon equation and its traveling wave solutions are well known and represent classical periodic, nonperiodic and kink solutions \cite{GrHarWin1,GrHarWin2}. For example, 
in the special case where $K_0=0$ and $K_1=0$, we obtain the following particular wave solution which is expressed in terms of Jacobi elliptic functions
\begin{equation}
\begin{split}
&\alpha(\sigma)=\arccos{\left(\mbox{cn}(\sigma,i)\right)},\qquad \sigma=x-t,\\
&\mu(\sigma)=\underline{D_1}\left[{\left(1-{\mbox{sn}^2(\sigma,i)\over \left(1+\mbox{dn}(\sigma,i)\right)^2}\right)\over   \left(1+{\mbox{sn}(\sigma,i)\over 1+\mbox{dn}(\sigma,i)}\right)^2}+{\left(1+{\mbox{sn}(\sigma,i)\over 1+\mbox{dn}(\sigma,i)}\right)^2\over \left(1-{\mbox{sn}^2(\sigma,i)\over \left(1+\mbox{dn}(\sigma,i)\right)^2}\right)}\right],\\
&\nu(\sigma)=\underline{D_1}\left[-{\left(1-{\mbox{sn}^2(\sigma,i)\over \left(1+\mbox{dn}(\sigma,i)\right)^2}\right)\over   \left(1+{\mbox{sn}(\sigma,i)\over 1+\mbox{dn}(\sigma,i)}\right)^2}+{\left(1+{\mbox{sn}(\sigma,i)\over 1+\mbox{dn}(\sigma,i)}\right)^2\over \left(1-{\mbox{sn}^2(\sigma,i)\over \left(1+\mbox{dn}(\sigma,i)\right)^2}\right)}\right],
\end{split}
\label{d3}
\end{equation}
where $\underline{D_1}$ is an arbitrary odd supernumber. Physically, this represents an elliptic traveling wave.

Another type of traveling wave solution is obtained for $\epsilon=-1$. A particular explicit solution of the reduced equations (\ref{res4}) takes the form
\begin{equation}
\begin{split}
&\Phi(x,t,\theta_1,\theta_2) = \arcsin{\left(\tanh{\sigma}\right)}+\theta_1 {\underline{D_1}\over \cosh{\sigma}} + \theta_2 \underline{D_1}\tanh{\sigma} - \theta_1 \theta_2 \tanh{\sigma},
\end{split}
\label{d5}
\end{equation}
where $\sigma=x+t$ and $\underline{D_1}$ is an arbitrary odd supernumber. This represents a bump function in $\theta_1$ direction and a kink in $\theta_2$ direction (i.e. in the corresponding odd components).

For the cases of subalgebras $\mathfrak{S}_6=\{P_x+\underline{\mu}Q_x\}$ and $\mathfrak{S}_{11}=\{P_t+\underline{\nu}Q_t\}$, the only non-zero solution which we obtain is
\begin{equation}
\Phi(x,t,\theta_1,\theta_2)=k\pi,\mbox{ where }k\in\mathbb{Z}.
\label{gian2}
\end{equation}

For subalgebra $\mathfrak{S}_7=\{P_t+\underline{\mu}Q_x\}$, we obtain the solutions
\begin{equation}
\Phi(x,t,\theta_1,\theta_2)=k\pi,
\label{gian1D}
\end{equation}
where $k\in\mathbb{Z}$ and
\begin{equation}
\Phi(x,t,\theta_1,\theta_2)=(k+\frac{1}{2})\pi+\theta_1\underline{\mu}\underline{\lambda_0}\psi(\sigma)+\theta_2\underline{\lambda_0}+(-1)^{k+1}(\theta_1-\underline{\mu}t)\theta_2,
\label{gian1E}
\end{equation}
where $k\in\mathbb{Z}$, $\underline{\lambda_0}$ is an odd supernumber and $\psi$ is an arbitrary even--valued function of $\sigma=x+\underline{\mu}\theta_1t$.

For the subalgebra $\mathfrak{S}_8=\{P_x+\varepsilon P_t+\underline{\mu}Q_x\}$, we were able to obtain an explicit solution only if we assume that $\lambda$ and $\eta$ are multiples of $\underline{\mu}$. 
Then the equation for $\alpha$ doesn't involve odd unknowns and can be solved in terms of elliptic functions. (We note that in this case the reduced equations become very similar to those for $\mathfrak{S}_4$, see Table \ref{sre}, but not identical -- they differ by $\mu \alpha_\sigma $ term in $\eta_{\sigma}+\underline{\mu}\alpha_{\sigma}-\lambda\cos{\alpha}=0$.) We find 
\begin{equation}
\Phi(x,t,\theta_1,\theta_2)=\alpha(\sigma)+(\theta_1-\varepsilon\underline{\mu}t)\eta(\sigma)+\theta_2\lambda(\sigma)+(\theta_1-\varepsilon\underline{\mu}t)\theta_2\beta(\sigma),
\label{ginv14}
\end{equation}
where $\sigma=\varepsilon x-t+\underline{\mu}t\theta_1$ and the  even--valued function $\alpha$ is given in terms of the Jacobi elliptic function
\begin{equation}
\alpha=\arcsin{\left[k\hspace{1mm}\mbox{sn}\left(\sqrt{-\varepsilon}\hspace{2mm}\sigma,k\right)\right]},
\label{ginv15}
\end{equation}
where the modulus $k$ is restricted by the relation $|k|<1$. The latter condition ensures that the elliptic solutions possess one real and one purely imaginary period  when restricted to real $\sigma$. The  even--valued function $\beta$ is given by
\begin{equation}
\beta=-k\hspace{1mm}\mbox{sn}\left(\sqrt{-\varepsilon}\hspace{2mm}\sigma,k\right).
\label{ginv16}
\end{equation}
The odd--valued function $\lambda$ is given by $\lambda=\underline{\mu}g(\sigma)$ where $g$ is an  even--valued function of $\sigma$ which obeys the linear ordinary differential equation
\begin{equation}
g_{\sigma\sigma}+(\tan{\alpha})g_{\sigma}-\varepsilon(\cos^2{\alpha})g+\varepsilon(\cos{\alpha})\alpha_{\sigma}=0,
\label{ginv17}
\end{equation}
and the odd--valued function $\eta$ is given by $\eta=\underline{\mu}f(\sigma)$ where the  even--valued function $f$ is given by
\begin{equation}
f=\frac{\varepsilon}{\cos{\alpha}}g_{\sigma}.
\label{ginv18}
\end{equation}

Subalgebra $\mathfrak{S}_{10}=\{P_x+\underline{\nu}Q_t\}$ leads to the solutions
\begin{equation}
\Phi(x,t,\theta_1,\theta_2)=k\pi,
\label{gian1F}
\end{equation}
where $k\in\mathbb{Z}$ and
\begin{equation}
\Phi(x,t,\theta_1,\theta_2)=(k+\frac{1}{2})\pi+\theta_2\underline{\nu}\underline{\lambda_0}\psi(\sigma)+\theta_1\underline{\lambda_0}+(-1)^{k}(\theta_2-\underline{\nu}x)\theta_1,
\label{gian1G}
\end{equation}
where $k\in\mathbb{Z}$, $\underline{\lambda_0}$ is an odd supernumber and $\psi$ is an arbitrary even--valued function of $\sigma$.

For the subalgebra $\mathfrak{S}_{12}=\{P_x+\varepsilon P_t+\underline{\nu}Q_t\}$, we find a solution similarly as in the case of $\mathfrak{S}_8$. We have
\begin{equation}
\Phi(x,t,\theta_1,\theta_2)=\alpha(\sigma)+(\theta_2-\underline{\nu}x)\eta(\sigma)+\theta_1\lambda(\sigma)+(\theta_2-\underline{\nu}x)\theta_1\beta(\sigma),
\label{ginv9}
\end{equation}
where $\sigma=t-\varepsilon x+\underline{\nu}x\theta_2$ and the  even--valued function $\alpha$ is given by
\begin{equation}
\alpha=\arcsin{\left[k\hspace{1mm}\mbox{sn}\left(\sqrt{-\varepsilon}\hspace{2mm}\sigma,k\right)\right]},
\label{ginv10}
\end{equation}
where the modulus $k$ is restricted by the relation $|k|<1$. The even--valued function $\beta$ is given by
\begin{equation}
\beta=k\hspace{1mm}\mbox{sn}\left(\sqrt{-\varepsilon}\hspace{2mm}\sigma,k\right).
\label{ginv11}
\end{equation}
The odd--valued function $\lambda$ is given by $\lambda=\underline{\nu}g(\sigma)$ where $g$ is an even--valued function of $\sigma$ which obeys the linear differential equation for $g$
\begin{equation}
g_{\sigma\sigma}+(\tan{\alpha})\alpha_{\sigma}g_{\sigma}-\varepsilon(\cos^2{\alpha})g-\varepsilon(\cos{\alpha})\alpha_{\sigma}=0,
\label{ginv12}
\end{equation}
and the odd--valued function $\eta$ is given by $\eta=\underline{\nu}f(\sigma)$ where the even--valued function $f$ is given by
\begin{equation}
f=-\frac{1}{\cos{\alpha}}g_{\sigma}.
\label{ginv13}
\end{equation}

Let us now turn our attention to those subalgebras whose invariants possess a non--standard structure. Such subalgebras are distinguished by the fact that each of them admits an invariant expressed in terms of an arbitrary function of the superspace variables, multiplied by an odd supernumber. Such invariants are nilpotent and this causes complications in the computation. This aspect can be illustrated by means of the following example. The subalgebra $\mathfrak{S}_5=\{\underline{\mu}Q_x\}$ generates the first of the two one--parameter group transformations described in equation (\ref{b2}). Its invariants are $t$, $\theta_2$, $\Phi$ and any quantity of the form
\begin{equation}
\tau=\underline{\mu}f\left(x,t,\theta_1,\theta_2,\Phi\right),
\label{nonstandard1}
\end{equation}
where $f$ is an arbitrary function which can be either even or odd--valued. It is an open question as to whether or not for a particular choice of function $f$ a substitution of these invariants into the SSG equation (\ref{symmie2}) can lead to a reduced system of equations expressible in terms of the invariants. It is clearly not possible for an arbitrary function $f$. For example, in the case when $\tau=\underline{\mu}x\theta_1$, the system (\ref{symmie2}) transforms into the equation
\begin{equation}
\underline{\mu}x\theta_2{\mathcal A}_{t\tau}+\underline{\mu}x{\mathcal A}_{\tau\theta_2}+\sin{\mathcal A}=0,
\label{nonstandard2}
\end{equation}
for the field
\begin{equation}
\Phi={\mathcal A}\left(t,\tau,\theta_2\right).
\label{nonstandard3}
\end{equation}
The presence of the variable $x$ in equation (\ref{nonstandard2})  clearly demonstrates that we do not obtain a reduced equation expressible in terms of the invariants.

On the other hand, if we would attempt the reduction with respect to all the vector fields $\mu Q_x, \mu \in \Lambda_{odd}$ we immediately find that such vector fields do not form a subalgebra and we have to reduce with respect to the subalgebra generated by $\{Q_x,P_x\}$. That leads to $\Phi(t,\theta_2)$ and substituting into equation (\ref{symmie2}) we find the reduction
\begin{displaymath}\sin{\Phi}=0,\end{displaymath}
which allows again only the trivial solution
\begin{equation}
 \Phi = k \pi, \qquad k \in \mathbb{Z}.
\end{equation}

The other five subalgebras having non-standard invariants display similar features, and we list below the invariants expressed in terms of an arbitrary function of the superspace variables for each case.

\noindent Subalgebra\hspace{7cm}Non-standard invariant\\\\
\noindent $\mathfrak{S}_9=\{\underline{\nu}Q_t\}$\hspace{7cm}$\underline{\nu}f\left(x,t,\theta_1,\theta_2,\Phi\right)$\\\\
\noindent $\mathfrak{S}_{13}=\{\underline{\mu}Q_x+\underline{\nu}Q_t\}$\hspace{6cm}$\underline{\mu}\underline{\nu}f\left(x,t,\theta_1,\theta_2,\Phi\right)$\\\\
\noindent $\mathfrak{S}_{14}=\{P_x+\underline{\mu}Q_x+\underline{\nu}Q_t\}$\hspace{5cm}$\underline{\mu}\underline{\nu}f\left(t,\theta_1,\theta_2,\Phi\right)$\\\\
\noindent $\mathfrak{S}_{15}=\{P_t+\underline{\mu}Q_x+\underline{\nu}Q_t\}$\hspace{5cm}$\underline{\mu}\underline{\nu}f\left(x,\theta_1,\theta_2,\Phi\right)$\\\\
\noindent $\mathfrak{S}_{16}=\{P_x+\varepsilon P_t+\underline{\mu}Q_x+\underline{\nu}Q_t\}$\hspace{4cm}$\underline{\mu}\underline{\nu}f\left(\theta_1,\theta_2,\Phi\right)$\\\\
where, in each case, $f$ is an arbitrary function of its arguments.

From a more general perspective, the problem of nonstandard invariants can be expressed as follows. We recall that the construction of invariant solutions is based on a proper (local) choice of canonical coordinates on the space of independent and dependent coordinates such that it ``straightens the flow'' of the generator of one--parametric subgroup, i.e. we get one coordinate corresponding to the group parameter and the remaining ones are invariant with respect to the flow. In order to perform the reduction we have to assume that the group parameter can be expressed as a function of original independent coordinates (i.e. the orbits of the subgroup action are of codimension one in the space of independent variables). We choose the proper number of invariant coordinates as our new dependent coordinates and interpret them as functions of the remaining ones and the group parameter.

Once the differential equation(s) possessing the symmetry is expressed in these canonical coordinates, due to its symmetry we are guaranteed that the group parameter coordinate drops out of the equation(s) and we have an equation(s) with one less independent variable. If this(these) equation(s) is(are) still a partial differential equation(s) too difficult to tackle directly, we can repeat the procedure and further reduce (provided of course that the reduced equation has some symmetries).

Now the source of the problem becomes clear: in the commutative case we can always locally straighten the flow of any nonvanishing vector field, as is well known from differential geometry. On the other hand, once we allow anticommuting variables, we are not always able to find such a coordinate transformation, as we have just seen -- although we have found the proper number of invariants, the transformation is obviously non--invertible. Therefore, we are led to the conclusion that in the case of anticommuting independent variables not all symmetry generators allow a symmetry reduction -- we have to restrict our attention only to those which can be written as a partial derivative with respect to even coordinate in some suitable coordinate system on the supermanifold $X \times U$ (of course, a possibility is not excluded that in some particular case a solution constructed out of nonstandard invariants may exist -- but its existence and the consistency of the reduction is not guaranteed).

\section{Conclusions}\label{concl}

In this paper, we have performed a group--theoretical analysis of the $(1+1)$--dimensional supersymmetric sine--Gordon model. This was accomplished using two different approaches. 

In the first one, the decomposition (\ref{b1}) of the bosonic superfield was substituted into the SSG equation and decomposed into a system of partial differential equations for the component fields. Next, we focused directly on the SSG equation expressed in terms of a bosonic superfield involving odd, anticommuting independent variables. 
In each case, we have determined a Lie (super)algebra of symmetries of the supersymmetric system and classified all of its one--dimensional subalgebras. In the case of the SSG equation (\ref{b7}), the superalgebra of symmetries was computed through the use of a generalized version of the prolongation method.

For the decomposed system (\ref{c1}), no odd symmetry generators were obtained and the Lie algebra of symmetries was just a realization of the Poincar\'{e} algebra in $(1+1)$ dimensions on the superspace. On the other hand, the Lie superalgebra of symmetries of the SSG equation (\ref{b7}) is the full super--Poincar\'{e} algebra in $(1+1)$ dimensions. 

Through the use of the SRM we have constructed exact analytic solutions of the SSG equation. The reductions in the component decomposition were found to be a special subset of reductions in superspace. Solutions included constant, algebraic, trigonometric, hyperbolic and doubly periodic solutions in terms of Jacobi elliptic functions. In some cases the reductions lead to systems of coupled ordinary differential equations whose full solution is unknown and we had to content ourselves with some particular explicit solutions or solutions expressed in terms of an arbitrary solution of given inhomogeneous linear ordinary differential equation (e.g. equation (\ref{ginv12})).

In the superspace formulation we have encountered one complication not present in the ordinary bosonic case; namely, not all generators allow the corresponding reduction. The reason for this is that there may be no canonical coordinates on the superspace straightening the flow of such vector field. Presently, we don't know about any simple criteria that would allow to immediately identify such problematic vector fields, i.e. without computation of invariants. 

We note that the groups of symmetries found were those one could guess at the beginning from the structure of the SSG equation. Unfortunately, that is very often the case with such an investigation -- an involved and lengthy computation is needed in order to exclude possibility of hidden, unexpected, symmetries but otherwise brings nothing new. 

An interesting open question is whether the supersymmetries can be somehow directly detected in the component form using the methods of symmetry analysis of differential equations (if one doesn't know that the equation was constructed to be supersymmetric, of course). That is, does supersymmetry demonstrate itself in some way e.g. as a contact or conditional symmetry? If such a detection was possible it might be systematically applied to find enhanced hidden supersymmetry in some models, which would be of significant practical importance.

Also, it would be of interest to apply the method of Section \ref{sssGe} to other physically relevant nonlinear supersymmetric models. For example, it should be possible to perform a similar analysis on a model with more supersymmetries, e.g. the $N=2$ supersymmetric sine--Gordon model. Given the computational complexities involved, namely the form of the prolongations defined by formulae (\ref{symmie7A}), it seems rather necessary to use computer algebra systems to deal with the expressions in these cases and we defer such investigation to future work.

\subsection*{Acknowledgments}

The research of AMG and AJH was supported by research grants from NSERC of Canada. The research of L\v{S} was supported by the research plan MSM6840770039 and the project LC527 of the Ministry of Education of the Czech Republic. AMG also thanks P. Exner and I. Jex for hospitality and the ``Doppler Institute'' project LC06002 of the Ministry of Education of the Czech Republic for support during his visit to the Faculty of Nuclear Sciences and Physical Engineering, Czech Technical University in Prague. We are also grateful to Ji\v{r}\'{\i} Tolar and Vojt\v ech \v St\v ep\'an for interesting discussions on the topic of this paper.

{}

\label{lastpage}
\end{document}